\documentstyle[12pt,axodraw,epsfig]{article}

\newcommand{\imag}{\Im {\rm m}} 
\newcommand{\real}{\Re {\rm e}}

\begin{document}

\mbox{ } \\[-1cm]
\mbox{ }\hfill DESY 00--105\\
\mbox{ }\hfill SNUTP 00--018\\
\mbox{ }\hfill \today\\

\begin{center}
{\Large\bf Measuring MSSM CP--violating Phases through Spin\\[-1mm] 
        Correlated Supersymmetric Tri--lepton Signatures\\[1mm]
        at the Tevatron}\\[1cm]
S.Y. Choi$^1$, M. Guchait$^2$, H.S. Song$^3$ and W.Y. Song$^3$
\end{center}

\bigskip 

\begin{center}
$^1${\it Department of Physics, Chonbuk National University, Chonju 561--756,
         Korea}\\[1mm]
$^2${\it Deutsches Elektronen--Synchrotron DESY, D-22603 Hamburg, 
         Germany}\\[1mm]
$^3${\it CTP and Department of Physics, 
         Seoul National University, Seoul 151-742, Korea}
\end{center}

\bigskip
\bigskip 

\begin{abstract}
We provide a detailed analysis of the supersymmetric tri--lepton signals 
for sparticle searches at the Tevatron in the minimal supersymmetric 
standard model with general CP--violating phases but without flavor  
mixing among sfermions of different generation. The stringent experimental 
constraints on the CP--violating phases 
from the electron and neutron electric dipole moments 
are included in the analysis for two exemplary 
scenarios of the SUSY parameters; one with decoupled first two generation 
sfermions and the 
other with non--decoupled sfermions.  In both scenarios, the production 
cross section and the branching fractions of the leptonic chargino and 
neutralino decays are sensitive to CP--violating
phases. The production--decay spin correlations lead to several 
non--trivial CP--even observables such as the lepton invariant mass 
distribution and the lepton angular distribution, 
and several interesting T--odd (CP--odd) momentum triple products.
The possibility of measuring the CP--violating phases directly through those 
T--odd observables is investigated in detail.
\end{abstract}
%


\newpage

\section{Introduction}
\label{sec:introduction}

Supersymmetry is the currently best motivated extension of the Standard Model
(SM) of particle physics which allows to stabilize the gauge hierarchy 
without getting into conflict with electroweak precision data. 
Among all possible supersymmetric theories, the Minimal Supersymmetric
Standard Model (MSSM) occupies a special position. It is
not only the simplest, i.e. most economical, potentially realistic
supersymmetric field theory, but it also has just the right particle
content to allow for the unification of all gauge interactions \cite{GUT}.
\\ 

The R--parity preserving MSSM \cite{NHK} as a softly--broken SUSY model 
contains in general more than one--hundred independent physical parameters 
including about forty--five CP--violating physical phases \cite{DS} in the 
Lagrangian.  The CP--violating phases, if they are large,
can give a significant impact on not only various CP--odd observables 
but also the CP--even quantities such as sparticle 
masses \cite{BK}, production cross sections, branching fractions \cite{Choi}, 
LSP relic density \cite{FO}, Higgs boson masses and couplings \cite{PW}, 
CP violation in the $B$ and $K$ systems \cite{BCKO}, and so on. \\

The most stringent (indirect) constraints \cite{MS} on the MSSM 
CP--violating phases come from the precise measurements of the electron 
and neutron electric dipole moments (EDM), but the indirect EDM 
constraints depend strongly on the assumptions taken in the 
analysis without any {\it a priori} justifications.  
Relaxing the rather stringent assumptions, several recent works \cite{IN,KO} 
have shown that the constraints could be evaded without suppressing 
the CP--violating phases of the theory. 
One option \cite{KO} is to make the first two generations of scalar fermions 
very heavy so that one--loop EDM constraints are automatically evaded
while keeping the third-generation sfermions relatively light to preserve
naturalness.  This case can be naturally explained by the 
so--called effective SUSY models \cite{DG} where de--couplings of the 
first and second generation sfermions are invoked to solve the SUSY 
FCNC and CP problems without spoiling naturalness. Another possibility 
is that various SUSY parameters are arranged \cite{IN} to lead to partial 
cancellations among their contributions to the electron and neutron EDMs 
without taking very large sfermion masses. Consequently, it is not yet clear 
at all whether the CP--violating phases of the MSSM are small or not. 
Once supersymmetric particles are discovered at colliders, it will be 
therefore of great importance to directly measure the CP--violating phases 
as well as the other real parameters of the supersymmetric Lagrangian.\\

In this paper we illustrate how the presence of the CP--violating phases
affects observables which can be measured in the classical reaction
$p\bar{p}\rightarrow\tilde{\chi}^\pm_1 \tilde{\chi}^0_2$ with subsequent 
decays $\tilde{\chi}^\pm_1\rightarrow
\tilde{\chi}^0_1\,\ell^\pm \bar{\nu}_\ell$ and $\tilde{\chi}^0_2\rightarrow
\tilde{\chi}^0_1\,\ell^{\prime +}\ell^{\prime -}$
$(\ell,\,\ell^\prime =e,\,\mu)$ when the lightest neutralino 
$\tilde{\chi}^0_1$ is the lightest supersymmetric particle (LSP).
Such tri--lepton signatures have been investigated in several CDF and D0 
analyses at the Tevatron based on the models with only real SUSY parameters. 
That is to say, most of the works \cite{MOST} have been 
done under the assumption that all the couplings are related at the 
grand unification or Planck scale and they are real. 
In the light of the possibility of large CP--violating phases, 
we investigate systematically the impact of the CP--violating phases on
the SUSY tri--lepton signatures at the Tevatron, including the
constraints on the phases imposed by the electron and neutron EDMs and 
taking into account the full spin/angular correlations between the 
production and the leptonic decays of the associated chargino 
and neutralino pair. \\

The paper is organized as follows. In Sect.~\ref{sec:mixing} we describe 
the chargino, neutralino and flavor--preserving sfermion mixing phenomena 
on the most general footing and identify all the relevant physical 
CP--violating phases. Section~\ref{sec: e and n EDM} is devoted to the 
detailed discussion of the constraints 
by the electron and neutron EDM measurements on the CP--violating phases.
In Sect.~\ref{sec:chargino and neutralino production} 
we present in detail the formalism to describe the associated 
production of the chargino $\tilde{\chi}^\pm_1$ and the neutralino
$\tilde{\chi}^0_2$ in $p\bar{p}$ collisions at the Tevatron;
production helicity amplitudes, chargino and neutralino mass spectra, 
and chargino and neutralino polarization vectors. 
Section~\ref{sec: chargino and neutralino decay} is devoted to 
the detailed description of the decay modes of the chargino 
$\tilde{\chi}^\pm_1$ and the neutralino $\tilde{\chi}^0_2$ and their
branching fractions. In Sect.~\ref{sec:correlation} after explaining 
the method of obtaining 
the fully spin--correlated distributions of the associated 
production and decays of the chargino and 
neutralino, we investigate in detail the impact of the CP--violating
phases on various physical observables such as the total rate of the
tri--lepton signatures, the dilepton invariant mass distributions,
the lepton angle distribution as
well as the CP--odd triple products of three proton and/or lepton
momenta. Finally, we summarize our findings and conclude in
Sect.~\ref{sec:conclusion}. \\

\section{Supersymmetric Flavor Conserving Mixing}
\label{sec:mixing}

The existence of the non--trivial MSSM CP--violating phases is due to 
SUSY breakdown, so that the CP--violating phases appear in the
soft--breaking parameters and the mixing among sparticles due to 
the electroweak gauge symmetry breaking. As a whole, the MSSM has
three well--known sources of CP violation. The first is 
related to the two Higgs--boson doublets present in the model as both the 
$\mu$ parameter in the superpotential and the soft breaking parameter $B$
can be complex. We denote the phases of $\mu$ and $B$ by $\Phi_{\mu}$ 
and $\Phi_{B}$, respectively.
Secondly, there are three more phases $\{\Phi_{1},\Phi_{2},
\Phi_{3}\}$ related to the complex U(1)$_Y$, SU(2)$_L$ 
and SU(3)$_C$ gauginos masses. Finally, most of the other 
CP--violating phases originate from the flavor sector of the MSSM
Lagrangian, either in the scalar soft mass matrices 
or in the trilinear matrices. \\

The sfermion mass matrices are Hermitian so that only off-diagonal 
terms can be complex, but the trilinear matrices are in general $3\times 3$ 
matrices allowing for the complex diagonal entries. The effects of 
the phases associated with the off-diagonal terms on experimental 
observables are strongly suppressed by the same mechanism required 
to suppress the flavor changing neutral current effects. 
Therefore, we neglect these flavor--changing CP--violating phases in 
the present work assuming that all the scalar soft mass matrices and 
trilinear parameters are flavor diagonal and the complex trilinear terms  
$A_f$ with its phase $\Phi_{A_f}$ are proportional to the corresponding 
fermion Yukawa couplings.
Consequently, the gaugino masses $M_1$, $M_2$, $M_3$ and the 
higgsino mass parameter $\mu$ as well as the trilinear parameters $A_f$
can be complex in the CP--noninvariant theories. 
However, reparameterising the fields one can take $M_2$ to be
real and positive without any loss of generality and all other parameter 
choices are related to the specific choice by an appropriate R 
transformation. \\

Neglecting flavor mixing among sfermions, the sfermion mass matrix
squared is given by
\begin{eqnarray}
 {\cal M}^2_{\tilde{f}}=
 \left(\begin{array}{cc}
       \tilde{m}^2_{\tilde{f}_L}+m^2_f+D_{f_L} & m^2_{\tilde{f} LR} \\[2mm]
       m^{2 *}_{\tilde{f} LR}  & \tilde{m}^2_{\tilde{f}_R}+m^2_f+
            D_{\bar{f}_R}
       \end{array}\right).
\label{eq:sfermion mass matrix}
\end{eqnarray}
where the mixing term $m^2_{\tilde{f} LR}$, and $D_{f_L}$ and
$D_{\bar{f}_R}$ are:  
\begin{eqnarray}
&& m^2_{\tilde{f} LR} = -m_f\left(A^*_f+\mu\tan\beta/\cot\beta \right)\ \  
        {\rm for} ~~f=d,e/u\,, \nonumber\\
&& D_{f_L}=m_Z^2\cos 2\beta \left(T^f_{3L}-Q_f\, s^2_W\right)\,,\qquad
   D_{\bar{f}_R}=m_Z^2\cos 2\beta\, Q_f\, s^2_W\,.
\end{eqnarray}
The first term of each diagonal element is the soft scalar mass term 
evaluated at the weak scale, and the second is the mass squared of 
the corresponding sfermion mass, and the last one is the so-called   
$D$ term of the MSSM superpotential. 
The trilinear term $A_f$ causing the left--right mixing is due to 
the soft--breaking Yukawa--type interaction,
and $\mu$ is the complex supersymmetric higgsino mass parameter 
and $\tan\beta$ is the ratio $v_2/v_1$
of the vacuum expectation values of the two neutral Higgs fields which
break the electroweak gauge symmetry. The sfermion mass eigenvalues
and eigenstates can be 
obtained by diagonalizing the above mass matrix with a unitary matrix
$U_{\tilde f}$ such that $U_{\tilde f}^\dagger\, {\cal M}^2_{\tilde{f}}\, 
U_{\tilde f} 
 = {\rm diag}(m^2_{\tilde{f}_1},m^2_{\tilde{f}_2})$. 
We parameterize $U_{\tilde f}$ as 
\begin{eqnarray}
U_{\tilde f}=\left(\begin{array}{cc}
    \cos\theta_f       &  -\sin\theta_f\,{\rm e}^{-i\phi_f} \\[1mm]
  \sin\theta_f\,{\rm e}^{i\phi_f}  & \cos\theta_f
          \end{array}\right),
\end{eqnarray}
where $\phi_f={\rm arg}[-m_f\,(A_f+\mu^*\tan\beta)]$, 
$0\leq\theta_f\leq\pi$ and $-\pi/2\leq\phi_f\leq\pi/2$ without
any loss of generality. \\

In the MSSM, the spin--1/2 supersymmetric partners of the $W^\pm$ boson and
the charged Higgs boson, $\tilde{W}^\pm$ and $\tilde{H}^\pm$,
respectively, mix to form the chargino mass eigenstates 
$\tilde{\chi}^\pm_{1,2}$. 
The chargino mass matrix dictating this mixing is given in the 
$\{\tilde{W}^-,\tilde{H}^-\}$ basis by
\begin{eqnarray}
{\cal M}_C=\left(\begin{array}{cc}
   M_2                   &    \sqrt{2}\,m_W\, c_\beta  \\[1mm]
 \sqrt{2}\,m_W\, s_\beta    &     |\mu|\,{\rm e}^{i\Phi_\mu}          
                 \end{array}\right),
\end{eqnarray}
which is built up by the fundamental SUSY parameters; the SU(2)$_L$ gaugino
mass $M_2$, the modulus and phase of $\mu$, $s_\beta\equiv\sin\beta$ and 
$c_\beta\equiv\cos\beta$.
Since the chargino mass matrix ${\cal M}_C$ is not symmetric, two
different unitary matrices acting on the left-- and right--chiral 
$(\tilde{W},\tilde{H})$ states are needed to diagonalize the
mass matrix ${\cal M}_C$:
\begin{eqnarray}
U_{L,R}\left(\begin{array}{cc}
             \tilde{W}^-   \\
             \tilde{H}^-
             \end{array}\right)_{L,R}
     = \left(\begin{array}{cc}
             \tilde{\chi}^-_1 \\
             \tilde{\chi}^-_1
             \end{array}\right)_{L,R},
\end{eqnarray}
so that $U_R{\cal M}_CU^\dagger_L={\rm diag}\,(m_{\tilde{\chi}^\pm_1},
m_{\tilde{\chi}^\pm_2})$ with the ordering of $m_{\tilde{\chi}^\pm_1}\leq
m_{\tilde{\chi}^\pm_2}$ as a convention. \\

The supersymmetric spin--1/2 partners of 
the neutral U(1)$_Y$ and SU(2)$_L$ gauge bosons, 
$\tilde{B}$ and $\tilde{W}^3$, respectively, mix  
with the supersymmetric fermionic partners of the neutral Higgs bosons, 
$\tilde{H}^0_1$ and $\tilde{H}^0_2$, to form mass eigenstates. 
The four physical mass eigenstates $\tilde{\chi}^0_i$ ($i=1$ to 4), called 
neutralinos, are obtained by diagonalizing the $4\times 4$ neutralino
mass matrix
\begin{eqnarray}
{\cal M}_N=\left(\begin{array}{cccc}
          |M_1|\,{\rm e}^{i\Phi_1}       &              0                     
 &   -m_Zc_\beta s_W  &   m_Zs_\beta s_W \\
           0                    &             M_2         
 &    m_Zc_\beta c_W  &  -m_Zs_\beta c_W \\[0.5mm]
     -m_Zc_\beta s_W  &   m_Zc_\beta s_W
 &         0                    &     -|\mu|\,{\rm e}^{i\Phi_\mu} \\ 
      m_Zs_\beta s_W  &  -m_Zs_\beta c_W
 &        -|\mu|\,{\rm e}^{i\Phi_\mu}      &              0
                \end{array}\right),
\label{eq:neutralino mass matrix}
\end{eqnarray}
where $s_W\equiv\sin\theta_W$ and $c_W\equiv\cos\theta_W$.  
Since the neutralino mass matrix ${\cal M}_N$ is a complex and symmetric
so that it can be diagonalized by a single unitary matrix $N$ as 
$ N^*{\cal M}_N N^\dagger = {\rm diag}\,(m_{\tilde{\chi}^0_1},
m_{\tilde{\chi}^0_2}, m_{\tilde{\chi}^0_3},m_{\tilde{\chi}^0_4})$ with the 
ordering of 
$m_{\tilde{\chi}^0_1}\leq m_{\tilde{\chi}^0_2}\leq m_{\tilde{\chi}^0_3}
\leq m_{\tilde{\chi}^0_4}$. 
On the other hand, the gluinos, the spin--1/2 partners of gluons,
do not mix among themselves and involve only a single complex 
phase $\Phi_3$. \\

To recapitulate, we have in total twelve non--trivial CP--violating phases;
two phases $\{\Phi_1,\Phi_3\}$ from the gaugino sector,
one phase $\Phi_\mu$ from the higgsino sector, and nine phases
$\Phi_{A_f}$ for the nine sfermion flavors.

\section{Electron and Neutron Electric Dipole Moments}
\label{sec: e and n EDM}

\begin{figure}
\begin{center}
\hbox to\textwidth{\hss\epsfig{file=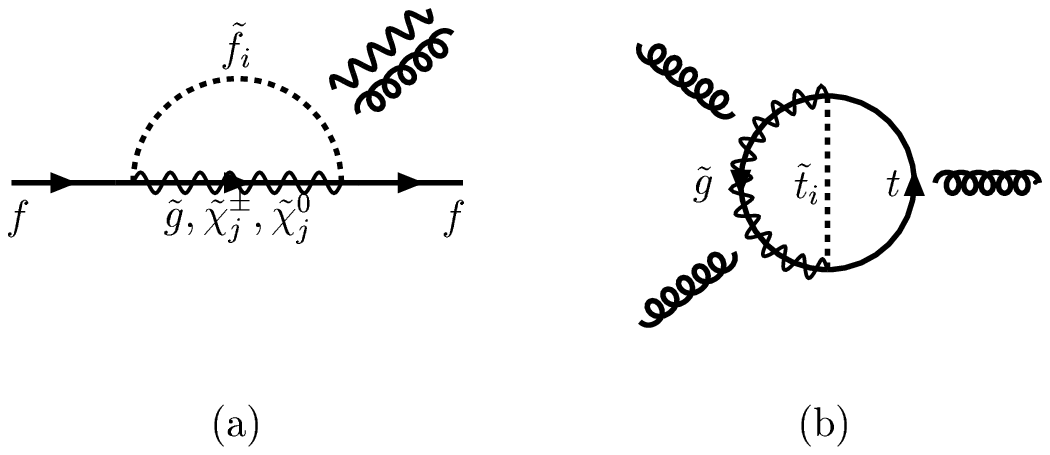,width=10cm,height=4cm}\hss}
\end{center}
\caption{\it The Feynman diagrams contributing to the fermion EDMs and CEDMs in 
         the MSSM; (a) one-loop chargino, neutralino and gluino diagrams
         and (b) two-loop top--Higgs or top quark--top squark--gluino diagrams.}
\label{fig:fig1}
\end{figure}

The electric dipole interaction of a spin--1/2 charged particle $f$ with an 
electromagnetic field is described in a model--independent way
by the 5--dimensional effective Lagrangian
\begin{eqnarray}
{\cal L}_{\rm EDM}=-\frac{i}{2}\, d_f\,\bar{f}\sigma^{\mu\nu}\gamma_5 f
                                  \, F_{\mu\nu}.
\end{eqnarray}
In theories with CP--violating interactions, the electric dipole moment 
$d_f$ receives contributions from loop diagrams. In the case of the 
electron EDM the chargino--sneutrino and neutralino--slepton loops contribute
whereas in the quark case the chargino--squark, neutralino--squark 
and gluino--squark loops are involved (See Fig.~\ref{fig:fig1}). 
Recently, one--loop chromoelectric dipole moment 
(CEDM) contributions have been studied in Ref.~\cite{IN} and found to be 
comparable to the EDM ones. In addition, there is a significant contribution 
to the neutron EDM through the so--called Weinberg's three--gluon 
dimension--six operator \cite{IN,glu}.  Therefore, for the CEDMs of quarks 
we include the chargino--squark, neutralino-squark and gluino--squark loops,
where the gluonic dimension--six operator gets contributions from the loop 
containing top (s)quark and gluinos. The quark CEDM is then
defined as the coefficient $\hat{d}_q$ in the 5-dimensional effective 
Lagrangian 
\begin{eqnarray}
{\cal L}_{\rm CEDM}=-\frac{i}{2}\,\hat{d}_q\,\bar{q}\sigma^{\mu\nu}
          \gamma_5 T^a q\, G^a_{\mu\nu}\,,
\end{eqnarray}
where the indices $a$ ($=1$ to 8) denote the gluons and $T^a$ are the 
SU(3)$_C$ generators. And, the gluonic dimension-six operator is given by
\begin{eqnarray}
{\cal L}_{\rm 3G}=-\frac{1}{6}\,d_G f_{abc} \, \epsilon^{\mu\nu\lambda\sigma}\, 
             G^a_{\mu\rho}G^{b\rho}_{\nu} G^c_{\lambda\sigma}\,.
\end{eqnarray}
The various SUSY contributions to the fermion EDM are 
illustrated  in Fig.~\ref{fig:fig1}; the diagram (a) denotes the chargino, 
neutralino, and gluino loops where the dashed line of each loop is for 
a scalar fermion mass eigenstate corresponding to the fermion, and 
the diagram (b) is for the effective 3--gluon operator. Both of
them contribute to the coefficients $d_f$ and $\hat{d}_q$ depending on 
whether a photon or a gluon is radiated
off the loop. These contributions can be calculated at the electroweak scale 
since a typical SUSY scale in most SUSY models is of the same order 
of magnitude 
as the electroweak scale. \\

In the following we enumerate all the SUSY contributions to the 
first--generation
fermion EDMs. The chargino--loop contributions are determined by
the Lagrangian 
\begin{eqnarray}
{\cal L}_{\tilde{\chi}^\pm f\tilde{f}^\prime} &=& \frac{e}{s_W} 
 \sum_{i=1}^2\sum_{j=1}^2
  \bigg\{\bar{e}\left[Y_e U^*_{Lj2}\delta_{i1}P_L 
                          - U^*_{Rj1}P_R\delta_{i1}\right]
         \tilde{\chi}^-_j\tilde{\nu}_i                    \nonumber \\ 
 && \hskip 1.8cm +\bar{d}\left[Y_d U^*_{Lj2} U_{\tilde{u}1i}P_L 
        +(Y_u U^*_{Rj2} U_{\tilde{u}2i} 
        - U^*_{Rj1} U_{\tilde{u}1i})P_R\right]
         \tilde{\chi}^-_j\tilde{u}_i                      \\ 
 && \hskip 1.8cm +\bar{u}\left[Y_u U_{Rj2} U_{\tilde{d}1i}P_L
        +(Y_d U_{Lj2} U_{\tilde{d}2i} 
        - U_{Lj1} U_{\tilde{d}1i})P_R\right]
         \tilde{\chi}^+_j\tilde{d}_i
  \bigg\}
 +{\rm H.c.},\nonumber
\end{eqnarray}
describing the $\tilde{\chi}^\pm$--$f$--$
\tilde{f}^\prime$ interactions for the first generation sfermions without 
flavor mixing. 
Here $P_{L,R}=(1\mp\gamma_5)/2$ and $\{Y_e,Y_d,Y_u\}$ are the
Yukawa couplings: 
\begin{eqnarray}
Y_e=\frac{m_e}{\sqrt{2}m_W\,c_\beta}\,,\ \
Y_d=\frac{m_d}{\sqrt{2}m_W\,c_\beta}\,,\ \
Y_u=\frac{m_u}{\sqrt{2}m_W\,s_\beta}\,.
\end{eqnarray}

One can readily find out that the interactions of the charginos with 
a fermion and a sfermion contribute to the EDM of the fermion $f$ as  
\begin{eqnarray}
d_f^{\tilde{\chi}^-} = -\frac{e\alpha}{4\pi s^2_W}
    \sum_{i=1}^2\sum_{j=1}^2
    \frac{m_{{\chi}^-_j}}{m^2_{\tilde{f}_i}}~
    \imag (\Delta^{f{\tilde{\chi}^-}}_{ij})
    \left[ Q_{\tilde{f}} \,
    B\!\left(\frac{m^2_{{\chi}^-_j}}{m^2_{\tilde{f}_i}}\right) 
  + (Q_f - Q_{\tilde{f}})\, 
    A\!\left(\frac{m^2_{{\chi}^-_j}}{m^2_{\tilde{f}_i}}\right) 
    \right],
\end{eqnarray}
where the dimensionless functions $A(r)$ and $B(r)$ are defined as
\begin{eqnarray}
A(r)=\frac{1}{2(1-r)^2}\left[3-r+2\,\left(\frac{\ln r}{1-r}\right)
\right]\,,
     \ \
B(r)=\frac{1}{2(1-r)^2}\left[1+r+2\,\left(\frac{r\ln r}{1-r}
\right)\right]\,,
\end{eqnarray}
and
\begin{eqnarray}
 \begin{array}{ll}
\Delta^{e{\tilde{\chi}^-}}_{ij} = 
      Y_e\, U^*_{Lj2} U_{Rj1}\delta_{i1}   &   \ \ 
     (Q_f = Q_e,\, Q_{\tilde{f}} = 0,\, 
      m_{\tilde{f}_i} = m_{\tilde{\nu}_e}),   \\[1mm]
\Delta^{u{\tilde{\chi}^-}}_{ij} = Y_u\, U_{Rj2} U^{\tilde{d}}_{1i}
           \left[ U^*_{Lj1} U^{\tilde{d}*}_{1i}
            - Y_d\, U^*_{Lj2} U^{\tilde{d}*}_{2i} \right]  &  \ \ 
     (Q_f = Q_u,\,  Q_{\tilde{f}} = Q_{\tilde{d}},\,
      m_{\tilde{f}_i} = m_{\tilde{u}_i}),       \\[1mm] 
\Delta^{d{\tilde{\chi}^-}}_{ij} = Y_d\, U_{Lj2} U^{\tilde{u}}_{1i}
           \left[ U^*_{Rj1} U^{\tilde{u}*}_{1i}
            - Y_u\, U^*_{Rj2} U^{\tilde{u}*}_{2i} \right]  &  \ \ 
     (Q_f = Q_d,\, Q_{\tilde{f}} = Q_{\tilde{u}},\,
      m_{\tilde{f}_i} = m_{\tilde{d}_i}).  
 \end{array}
\end{eqnarray}
On the other hand, the chargino contributions to the CEDM of the quark $q$ 
are given by 
\begin{eqnarray}
\hat{d}_q^{\tilde{\chi}^-} = -\frac{g_s\alpha}{4\pi s^2_W}
    \sum_{i=1}^2\sum_{j=1}^2
    \frac{m_{{\chi}^-_j}}{m^2_{\tilde{q}_i}}~
    \imag (\Delta^{q{\tilde{\chi}^-}}_{ij})\, 
   B\!\left(\frac{m^2_{{\chi}^-_j}}{m^2_{\tilde{q}_i}}\right). 
\label{eq:chargino contribution}
\end{eqnarray}

Secondly, the neutralino--loop contributions to the EDM of a fermion
$f$ are determined by the Lagrangian describing the most general 
$\tilde{\chi}^0$--$f$--$\tilde{f}$ interactions,  
\begin{eqnarray}
{\cal L}_{\tilde{\chi}^0 f\tilde{f}} = -\frac{e}{\sqrt{2}s_W}
 \sum_{i=1}^2\sum_{j=1}^4
  \bar{f}\left[B^{f_L}_{ij} P_L + B^{f_R}_{ij} P_R\right]
         \tilde{\chi}^0_j\tilde{f}_i
 +{\rm H.c.},
\end{eqnarray}
where the couplings $B^{f_L}_{ij}$ and $B^{f_R}_{ij}$ are 
given by
\begin{eqnarray}
&& B^{f_L}_{ij}=\sqrt{2}Y_f N^*_{jh} U_{\tilde{f}1i}
             -2Q_f N^*_{j1}\tan\theta_W U_{\tilde{f}2i}\,, \nonumber\\[0.5mm]
&& B^{f_R}_{ij}=2\left[T^f_3 N_{j2} 
              + (Q_f-T^f_3)\tan\theta_W N_{j1}\right] U_{\tilde{f}1i}  
               + \sqrt{2}Y_f N_{jh} U_{\tilde{f}2i}\,,  
\end{eqnarray}
with $h=3$ for $f$ = $d,e$ and $h$ = 4 for $f$ = $u$, respectively. 
The neutralino contributions to the EDM of a fermion $f$ is then 
given by 
\begin{eqnarray}
 d_f^{\tilde{\chi}^0} =\frac{e\alpha}{8\pi s^2_W} Q_f
   \sum_{i=1}^2\sum_{j=1}^4
   \left\{\frac{m_{\tilde{\chi}^0_j}}{m^2_{\tilde{f}_i}}\,
   \imag\left[B^{fL}_{ij} B^{fR*}_{ij}\right]
   B\!\left(\frac{m^2_{\tilde{\chi}^\pm_j}}{m^2_{\tilde{f}_i}}\right)
   \right\},
\label{eq:neutralino contribution}
\end{eqnarray}
and the neutralino contributions to the CEDM $\hat{d}_q^{\tilde{\chi}^0}$
of the quark $q$ can be obtained by simply replacing $e$ by $g_s$ and 
$Q_f$ by unity in eq.~(\ref{eq:neutralino contribution}). \\

Thirdly, the gluino--loop contributions to the fermion EDMs are determined by
the Lagrangian describing the most general $\tilde{g}$--$q$--$\tilde{q}$ 
interactions  
\begin{eqnarray}
{\cal L}_{\tilde{g}q\tilde{q}} = \sqrt{2}g_s (T^a)_{lm}
       \sum_{i=u,d}\left(
       e^{-i\Phi_3 /2}\,\bar{q}^l_i P_L\tilde{g}^a\tilde{q}^m_{iR}
     + e^{i\Phi_3 /2}\,\bar{q}^l_i P_R\tilde{g}^a\tilde{q}^m_{iL} \right)
 +{\rm H.c.}\,,
\end{eqnarray}
where the indices $l$ and $m$ ($=1$ to $3$) are for the color of 
quarks and squarks, the index $a$ ($=1$ to 8) is for the color of
a gluino $\tilde{g}^a$, and the 3$\times$3 matrix $T_a$ is  
a SU(3)$_C$ generator. The gluino contributions to the coefficients 
$d_q$ and $\hat{d}_q$ are given by
\begin{eqnarray}
&& d_q^{\tilde{g}} = -\frac{2e\alpha_s}{3\pi} Q_q\sum_{i=1}^2
  \frac{m^2_{g}}{m^2_{\tilde{q}_i}}
 \imag(e^{-i\Phi_3} U_{\tilde{q}2i} U^*_{\tilde{q}1i})\, B(x)\,,
 \nonumber\\
&& \hat{d}_q^{\tilde{g}} = \frac{g_s\alpha_s}{4\pi}\sum_{i=1}^2
  \frac{m^2_{g}}{m^2_{\tilde{q}_i}}
  \imag(e^{-i\Phi_3} U_{\tilde{q}2i} U^*_{\tilde{q}1i})\, C(x)\,,
\label{eq:gluino contribution}
\end{eqnarray}
with the dimensionless function $C(x)$ defined as
\begin{eqnarray}
C(r)= - 3A(r) + \frac{B(r)}{3}\,.
\end{eqnarray}\\[-4mm]

Finally,\, the leading nontrivial MSSM contribution to the CP--odd 
three--gluon term $d_{G}$ is given by a two--loop diagram involving 
the top, top squarks and gluinos \cite{IN}:
\begin{eqnarray}
d_G &=& \frac{3g_s\alpha_s^2}{16\pi^2} 
  \frac{m_t(m^2_{\tilde{t}_2}-m^2_{\tilde{t}_1})}{m^5_{\tilde{g}}}\,
  \imag(e^{-i\Phi_3} U^{\tilde{t}}_{22} U^{\tilde{t}*}_{12})\, H(x_1,x_2,x_3)
 \,,
\label{eq:3-gluon contribution}
\end{eqnarray}
where $x_1=m^2_{\tilde{t}_1}/m^2_{\tilde{g}}$,
 $x_2=m^2_{\tilde{t}_2}/m^2_{\tilde{g}}$, $x_3=m^2_{t^2}/m^2_{\tilde{g}}$,
and the two--loop function $H$ is given by the three--fold integral
\begin{eqnarray}
H(z_1,z_2,z_3)=\frac{1}{2} \int_0^1 {\rm d}x\int_0^1 {\rm d}u \int_0^1 
               {\rm d}y\, x (1-x) u\, \frac{N_1 N_2}{D^4},
\end{eqnarray}
where the explicit forms of the $N_1$, $N_2$ and $D$ are given by
\begin{eqnarray*}
N_1&=&u(1-x)+z_3 x (1-x) (1-u)-2 ux [z_1 y+z_2(1-y)]\,,\\
N_2&=&(1-x)^2(1-u)^2+u^2-\frac{1}{9} x^2(1-u)^2\,,  \\
D&=&u(1-x)+z_3 x (1-x) (1-u)+ux[z_1 y+z_2 (1-y)]\,.
\end{eqnarray*}\\[-4mm]

Having defined the contributions from the individual Feynman diagrams,
the total EDM of electron is given simply by the sum of
the chargino and neutralino contributions:
\begin{equation}
d_e = d_e^{\tilde{\chi}^{+}} + d_{e}^{\tilde{\chi}^{0}}\,.
\end{equation}
In principle, the Kobayashi--Maskawa (KM) CP phase in the SM can contribute 
to the electron EDM, but it turns out to be effective
only at the three--loop level so that the contribution is too small 
to be considered.\\

On the other hand, the EDM of the neutron, a composite particles of quarks 
and gluons, can be obtained from the EDMs of the constituent quarks and 
gluons. The quark EDM evaluated at the electro--weak scale must be evolved 
down to the hadronic scale via renormalization group equation (RGE). 
Generally, the prescription based on  the effective chiral quark 
theory given in Ref.~\cite{Manohar}, are used to calculate the
neutron EDM from the quark EDMs.
According to the model, the neutron EDM is given by
\begin{eqnarray}
d_n=\frac{4}{3}\, d_d-\frac{1}{3}\, d_u.
\end{eqnarray}
In order to evaluate the numerical value of each individual quark
EDM from the effects of the EDM operators given in
eqs.~(\ref{eq:chargino contribution}), (\ref{eq:neutralino contribution}),
(\ref{eq:gluino contribution}) and (\ref{eq:3-gluon contribution})
we use the so--called ``naive dimensional analysis" as proposed
in Ref.~\cite{Manohar}.
The explicit form of the quark EDMs are then given by the contributions of 
all quark and gluon operators (to leading order in $\alpha_{s}$) with the
proper dimensional re--scalings as 
\begin{eqnarray}
d_q = \eta^{E}
  ( d_q^{\tilde{\chi}^{+}} + d_q^{\tilde{\chi}^{0}}
  + d_{q}^{\tilde{g}} )
+ \eta^{C} \frac{e}{4\pi}
  ( \hat{d}_{q}^{\tilde{\chi}^{+}}
  + \hat{d}_{q}^{\tilde{\chi}^{0}}
  + \hat{d}_{q}^{\tilde{g}} )
+ \eta^{G}\, \frac{e\Lambda_{SB}}{4\pi} d_G\,,
\label{chiral-quark-edm}
\end{eqnarray}
where $\eta^{E}$, $\eta^{C}$, and $\eta^{G}$ are the QCD correction factors
due to the renormalization group equations, whereas $\Lambda_{SB}$ is 
the scale of chiral symmetry breaking in QCD; we use 
$\eta^{E}=1.53$ \cite{Arnowitt}, $\eta^{C} \simeq \eta^{G} \simeq 3.4$  
\cite{IN}, and $\Lambda_{SB} \simeq 1.19$~GeV \cite{Manohar}.\\

Although flavor mixing is neglected, there are still sixteen 
real parameters and seven CP--violating phases: $\{\tan\beta,\,|M_1|,\,
M_2,\,|M_3|,\,|\mu|,\, m_{\tilde{f}_L},\,m_{\tilde{f}_R},\, |A_f|\}$
and $\{\Phi_1,\Phi_3,\Phi_\mu,\Phi_{A_f}\}$ for $f=e,u,d,t$ where
the SU(2) relation $m_{\tilde{u}_L}=m_{\tilde{d}_L}$ is taken
into account.
So, any reasonable quantitative understanding of the general features of the 
SUSY contributions to the electron and neutron EDMs requires us to make 
a few appropriate assumptions on the 
SUSY parameters without spoiling their 
qualitative aspects; (i) we take a universal soft--breaking 
selectron mass $m_{\tilde{e}}$ and squark mass $m_{\tilde{q}}$ for 
the first--generation left-- and right--handed selectrons and squarks,
respectively, and $m_{\tilde{t}}$ for the left-- and
right--handed top squarks; (ii) the gaugino mass unification condition 
is assumed only for the modulus of the U(1)$_Y$ and SU(2)$_L$
gaugino mass parameters, i.e.
$|M_1|=\frac{5}{3}\tan^2\theta_W\,M_2\approx 0.5\,M_2$ and
the absolute value of the gluino mass, $|M_3|$, is taken to be 500 GeV; 
(iii) we take $|A_f|$ to be 1 TeV for any sfermion $\tilde f$ 
such that the EDM constraints are satisfied.
We should be very careful in choosing the value of 
$\tan\beta$ which has very significant effect to many SUSY processes.
According to the recent calculation for the Barr--Zee--type
two--loop contributions to the fermion EDMs by Chang, Keung and Pilaftsis 
\cite{CKP}, the bottom squark contributions are very much enhanced 
for large $\tan\beta$ \cite{BCDP} so that their contributions cannot be 
neglected. Therefore, for large $\tan\beta$ we are forced to 
introduce additional CP--violating phases in our analysis related with 
the sparticles of the third generation.  Moreover, a large $\tan\beta$ 
yields a large tau slepton left--right mixing and the Higgs--exchange 
diagrams in the decays of the neutralinos into tau or bottom pairs cannot 
be neglected due to the enhanced Yukawa couplings of third generation. 
On the contrary, the value of $\tan\beta$ less than about 2 
has been already ruled out by negative results in the Higgs search 
experiments at LEP II \cite{HIGGS} based on the minimal 
supergravity scenario with real couplings.
However, this experimental constraint on $\tan\beta$ may be
loosened in CP noninvariant theories.
Putting off the detailed investigation related with the 
$\tan\beta$ dependence of the EDMs, the associated production of chargino 
and neutralino, and their branching ratios, we simply 
take $\tan\beta=3$ in the present analysis, for which the tau-slepton 
contributions can be treated on the same footing as the other slepton 
contributions. \\

\begin{figure}
\begin{center}
\hbox to\textwidth{\hss\epsfig{file=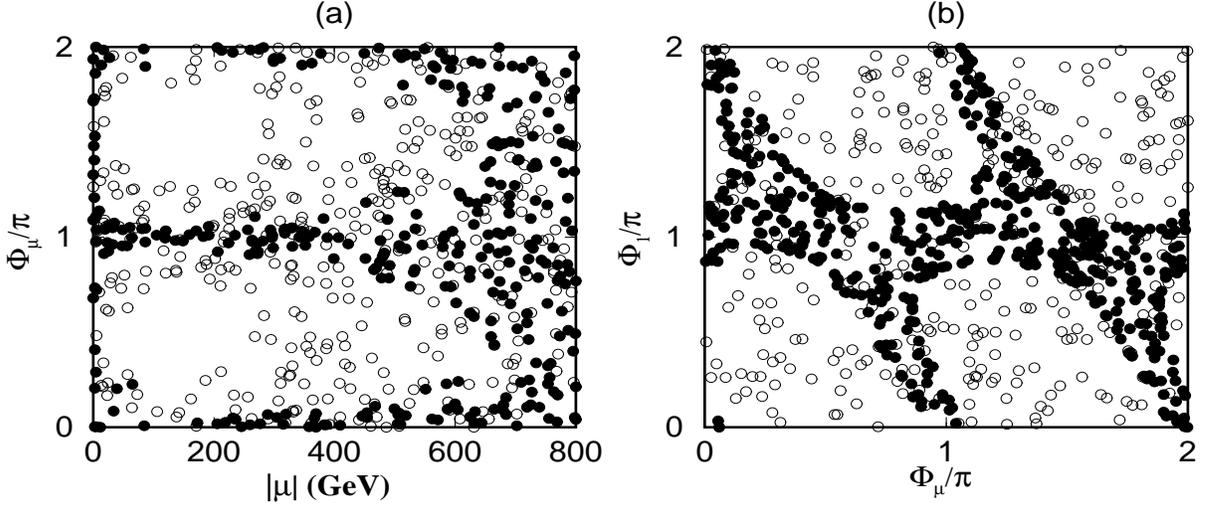,width=16cm,height=7cm}\hss}
\end{center}
\caption{\it (a) the allowed range of the CP--violating phase $\Phi_\mu$ 
         versus the higgsino mass parameter $|\mu|$ and (b) the allowed 
         region of $\{\Phi_\mu,\Phi_1\}$ against the electron (filled circles) 
         and neutron (hollow circles) EDM constraints in the scenario
         ${\cal S}2$.
         The trilinear parameter $|A_e|$ is taken to be 1 TeV and its phase
         $\Phi_{A_e}$ is scanned over the full allowed range.}
\label{fig:fig2}
\end{figure}

With these assumptions on the SUSY parameters given in the previous 
paragraph, the electron and neutron EDMs are determined by just five real 
parameters and seven remaining CP--violating phases:
\begin{eqnarray}
\{M_2,\,|\mu|,\,m_{\tilde{e}},\,m_{\tilde{q}},\,m_{\tilde{t}}\,\, ;
\,\,\Phi_1,\,\Phi_\mu,\,\Phi_3,\, \Phi_{A_e},\,\Phi_{A_q},\, \Phi_{A_t}\}\,,
\end{eqnarray}
with $q=u,\, d$. Let us take two extreme scenarios for allowing
relatively large CP--violating phases without violating the EDM
constraints. Firstly, based on the so-called effective SUSY model \cite{DG}
we decouple the first and second generation sfermions by rendering 
them very heavy without violating naturalness by maintaining the
third generation sfermions to be light and
taking $M_2$ and $|\mu|$ relatively small: 
\begin{eqnarray}
{\cal S}1:\ \ \{M_2=100\,{\rm GeV},\,\,|\mu|=200\,{\rm GeV},\,\, 
m_{\tilde{f}}=10\,{\rm TeV},\,\, m_{\tilde{t}}=500\,{\rm GeV}\}\,,
\end{eqnarray}
In this scenario, the sfermion masses   
$m_{\tilde{f}}$ $(f=e,u,d)$ are large enough to suppress  
the sfermion contributions to the electron and neutron EDMs completely. 
As a result, the present EDM bounds \cite{eEDM} on the electron and neutron 
EDMs 
\begin{eqnarray}
|d_e|\leq 4.3\times 10^{-27}\,e\cdot{\rm cm}\,,\qquad
|d_n|\leq 1.1\times 10^{-25}\,e\cdot{\rm cm}\,,
\end{eqnarray}
put no constraints at all on the CP--violating phases 
$\{\Phi_{1,3},\,\Phi_\mu\}$. Secondly, for the sake of generality,
we consider a scenario with relatively small universal soft--breaking 
sfermion masses but with a large value of $|\mu|$ as
\begin{eqnarray}
{\cal S}2:\ \ \{M_2=100\,{\rm GeV},\,\, |\mu|=700\,{\rm GeV},\,\,
m_{\tilde{e}}=200\,{\rm GeV},\,\, m_{\tilde{u},\tilde{d},\,
\tilde{t}}=500\,{\rm GeV}\}.
\end{eqnarray}
This scenario of small sfermion masses requires some cancellations among 
the different contributions so as to escape the electron and neutron EDM 
constraints. The degree of cancellations depends on the moduli of the  
SUSY parameters such as $|\mu|$, $|M_1|$ and $M_2$. The reason for 
taking a large $|\mu|$ ($=700$ GeV) is because it
allows a relatively large region for the CP phases $\Phi_\mu$ and 
$\Phi_1$ for any values of the phase $\Phi_{A_f}$ between 0 to $2\pi$.
The fact that the allowed region of two phases increases with $|\mu|$ 
has been pointed out in several previous works \cite{IN}. 
In Fig.~\ref{fig:fig2}(a), the allowed range of $\Phi_\mu$ versus 
$|\mu|$ at 95\% confidence level is displayed with the other real
parameters of ${\cal S}2$ and the other phases sampled 
randomly within their allowed ranges. The overall trend is 
that for larger $|\mu|$ it is much easier to satisfy the EDM limits and 
any $|\mu|\geq 650$ GeV allows the full range of $\Phi_\mu$. 
Except for small $|\mu|$, the neutron EDM constraints are less stringent
than that of the electron. Fig.~\ref{fig:fig2}(b) shows the allowed region 
at 95\% confidence level for the
phases $\Phi_\mu$ and $\Phi_1$ in the scenario ${\cal S}2$.
Note that the electron EDM constraints allow the full range of $\Phi_\mu$ only 
for $\Phi_1$  around $\pi$ while the neutron EDM constraints 
give no restrictions to $\Phi_\mu$ and $\Phi_1$ in the scenario ${\cal S}2$.
It is also worthwhile to note that the allowed region of 
two phases can be enlarged by taking large values of $M_2$ while 
keeping $|\mu|$ relatively small. From the above discussion, it
is clear that allowing large CP violating phases against the stringent
electron and neutron EDM constraints needs some sort of ``fine tuning"
among the SUSY parameters.\\

On the other hand, the spin--correlated production and decays 
of the associated chargino and neutralino pair in $p\bar{p}$ collisions
is mainly dependent on the flavor--independent CP--violating phases 
$\Phi_\mu$ and $\Phi_1$ but nearly independent of the flavor--dependent
phases $\Phi_{A_f}$ and the phase $\Phi_3$ of the gluino mass.
In the following numerical analysis, Fig.~\ref{fig:fig2}(b) will be 
taken to be the basic 
platform for all the contour plots of the production and total cross sections 
of the correlated process, the branching fractions of the chargino and
neutralino decays, and the CP--odd (T--odd) triple momentum products
in the scenario ${\cal S}2$. 
Since the neutron EDM constraints on the CP--violating phases are weaker 
than the electron EDM ones in both scenarios ${\cal S}1$ and ${\cal S}2$ 
we include only the latter constraints on the CP--violating phases in 
our numerical analysis.

\section{Production of a Chargino and Neutralino Pair}
\label{sec:chargino and neutralino production}

The production of $\tilde{\chi}^\pm_1\tilde{\chi}^0_2$, followed by the 
subsequent decays $\tilde{\chi}^\pm_1\rightarrow\tilde{\chi}^0_1\,\ell^\pm
\nu_\ell$ and $\tilde{\chi}^0_2\rightarrow\tilde{\chi}^0_1\,\ell^{\prime +}
\ell^{\prime -}$, is a main source of three charged leptons 
($e$ or $\mu$) and $\not\!\!{E}_T$ in $p\bar{p}$ collisions, called 
tri--lepton events. Since the tri--lepton signal suffers by very tiny SM 
backgrounds, it is, therefore, considered to be one of the most promising
channels by which low energy SUSY can be discovered in hadron colliders,
especially at the Tevatron.
The cross section of $\tilde{\chi}^\pm_1\tilde{\chi}^0_2$ is determined 
mainly by the chargino and neutralino masses and the overall efficiency
for detecting the three final--state 
leptons is very sensitive to the mass splittings 
between $\{\tilde{\chi}^\pm_1,\tilde{\chi}^0_2\}$ and $\tilde{\chi}^0_1$
as well as the corresponding branching fractions.
In the following subsection, we investigate the dependence of chargino 
and neutralino mass spectra on the CP--violating phases $\Phi_\mu$ and 
$\Phi_1$ in the two scenarios ${\cal S}1$ and ${\cal S}2$. 
Then, we present the helicity amplitudes for the parton--level process 
$d\bar{u}\rightarrow \tilde{\chi}^-_i\tilde{\chi}^0_j$, allowing us to 
construct the production cross section and to investigate the chargino and 
neutralino polarizations.

\subsection{Chargino and neutralino masses}
\label{subsec:4.1}

The chargino and neutralino masses are obtained by diagonalising 
the mass matrices ${\cal M}_C$ and ${\cal M}_N$, respectively.
Unlike the neutralino masses, the chargino masses are independent
of the modulus and phase of the U(1)$_Y$ gaugino mass, $|M_1|$ and $\Phi_1$. 
Since, in both scenarios ${\cal S}1$ and ${\cal S}2$, the higgsino mass 
parameter $|\mu|$ is larger than 
the gaugino mass parameters $M_2$ and $|M_1|$, the lightest mass
eigenstates are gaugino--like and the heaviest states are higgsino--like.
This feature is expected to be more prominent in the scenario ${\cal S}2$ 
with the larger value of $|\mu|$.\\

\begin{figure}
\begin{center}
\hbox to\textwidth{\hss\epsfig{file=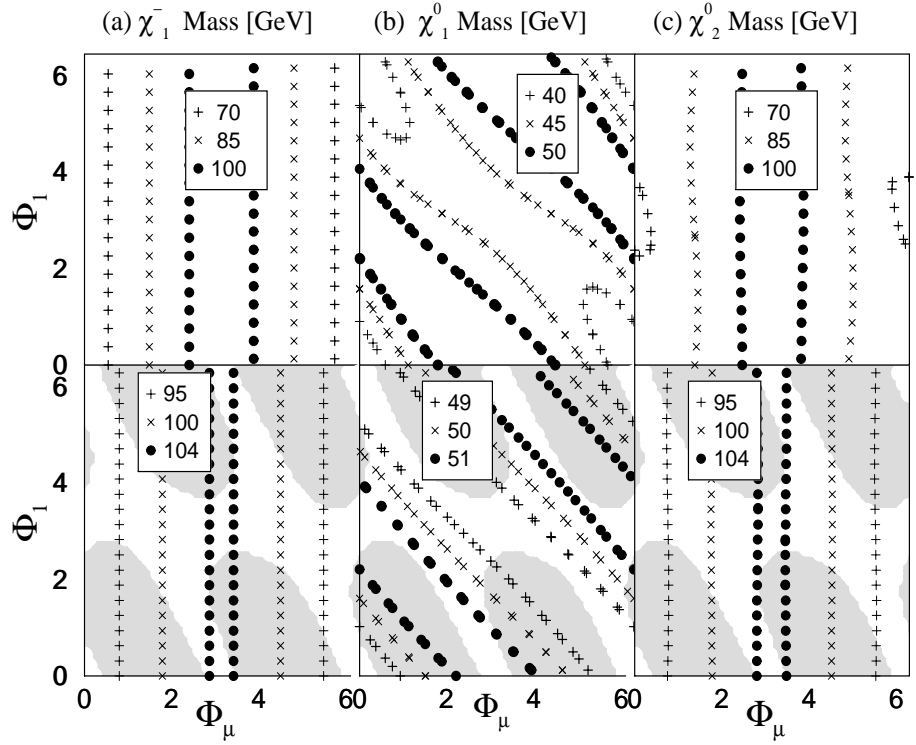,width=12cm,height=10cm}\hss}
\end{center}
\caption{\it The chargino and neutralino masses - (a) $m_{\tilde{\chi}^-_1}$, 
         (b) $m_{\tilde{\chi}^0_1}$ and (c) $m_{\tilde{\chi}^0_2}$ -  on 
         the $\{\Phi_\mu,\Phi_1\}$ plane in ${\cal S}1$ (upper frames) 
         and ${\cal S}2$ (lower frames).}
\label{fig:fig3}
\end{figure}

Figure~\ref{fig:fig3} shows the mass spectrum of the lightest chargino 
$\tilde{\chi}^\pm_1$ and the neutralinos $\tilde{\chi}^0_{1,2}$ 
on the $\{\Phi_\mu,\Phi_1\}$ plane for the scenarios ${\cal S}1$ and 
${\cal S}2$. 
Each shaded area of the lower figures is excluded  
by the electron EDM constraints.  
Except for the region of $\Phi_\mu=0,\,2\pi$, the masses 
$m_{\tilde{\chi}^\pm_1}$ and $m_{\tilde{\chi}^0_2}$ are very similar
in size and independent of $\Phi_1$ in both ${\cal S}1$ and ${\cal S}2$
while $m_{\tilde{\chi}^0_1}$ exhibits a strongly correlated
dependence on the CP-violating phases. 
The masses $m_{\tilde{\chi}^\pm_1}$ and $m_{\tilde{\chi}^0_2}$
increase as $\Phi_\mu$ approaches $\pi$, while $m_{\tilde{\chi}^0_1}$
becomes maximal at certain non--trivial values of $\Phi_\mu$ and $\Phi_1$ in 
${\cal S}1$. This implies that $m_{\tilde{\chi}^0_1}$ is 
strongly affected by a small value of $|\mu|$, while $m_{\tilde{\chi}^\pm_1}$ 
and $m_{\tilde{\chi}^0_2}$ are essentially determined by the SU(2)$_L$
gaugino mass $M_2$. The mass $m_{\tilde{\chi}^0_1}$
becomes smaller as the CP phases $\Phi_1$ and $\Phi_\mu$ approach the
off--diagonal line on the plane, implying that the
mass is a function of the sum $\Phi_\mu+\Phi_1$ of two CP phases to a very 
good approximation. Note that  the masses $m_{\tilde{\chi}^\pm_1}$ and 
$m_{\tilde{\chi}^0_{1,2}}$ are more sensitive to the 
phases in ${\cal S}1$ because $|\mu|$ is comparable with $M_2$ and $m_Z$ 
in size in this scenario.\\

\begin{figure}
\begin{center}
\hbox to\textwidth{\hss\epsfig{file=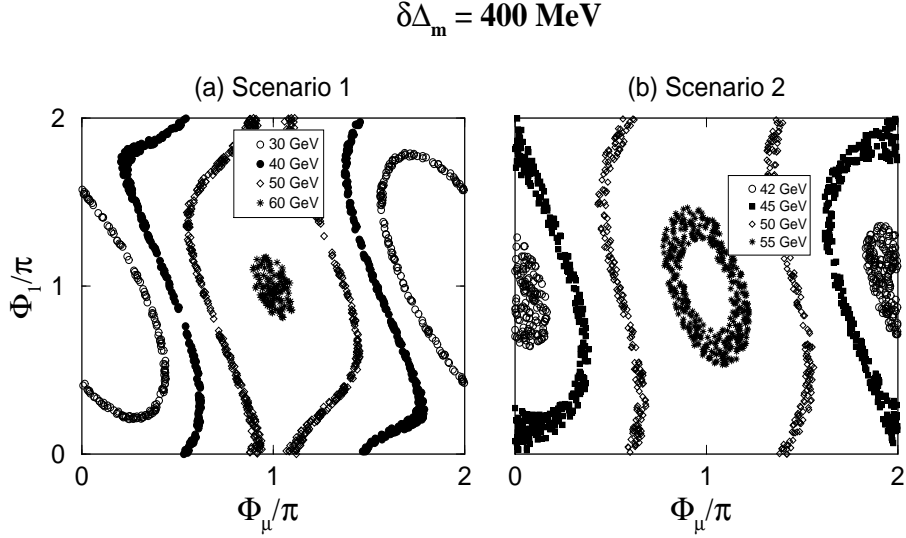,width=12cm,height=7.5cm}\hss}
\end{center}
\caption{\it The constraints on $\{\Phi_\mu, \Phi_1\}$ by the measurements of 
         the mass difference 
         $\Delta_m=m_{\tilde{\chi}^0_2}-m_{\tilde{\chi}^0_1}$ for four
         different mean values of the mass difference $\Delta_m$
         with the assumed uncertainty of 400 MeV in (a) ${\cal S}1$ and
         (b) ${\cal S}2$.}
\label{fig:fig4}
\end{figure}

Although the chargino and neutralino masses cannot be separately
measured at the Tevatron, the mass difference $\Delta_m\equiv
m_{\tilde{\chi}^0_2}-m_{\tilde{\chi}^0_1}$ can be determined with a 
good precision \cite{NY} by measuring the end points of the invariant mass 
distribution of the same flavor but opposite sign dileptons from the 
neutralino decay. The precision of determining the mass difference is 
sensitive to the event rate and dependent on experimental capabilities. 
Deferring the discussion on these aspects to 
Sect.~\ref{sec:chargino and neutralino production} we illustrate what 
information on the CP--violating phases the $\Delta_m$ measurement 
can provide us with. For that purpose, we assume the uncertainty for 
the mass difference to be 400 MeV and exhibit in Fig.~\ref{fig:fig4} 
the allowed area of the CP--violating 
phases $\{\Phi_\mu,\Phi_1\}$ for the SUSY parameters of (a) ${\cal S}1$ 
and (b) ${\cal S}2$. The figure clearly shows that the mass difference is 
a very sensitive probe of the CP--violating phases if the other real
parameters are known; the sensitivity of
the mass difference to the phases are enhanced when the gaugino and higgsino 
mass parameters are comparable to $m_Z$ in size.

\subsection{Parton--level production helicity amplitudes}
\label{subsec:4.2}

Although we are mainly interested in the production process 
$d\bar{u}\rightarrow\tilde{\chi}^-_1\tilde{\chi}^0_2$, we discuss
in this section the associated production of any chargino and
neutralino pair $d\bar{u}\rightarrow\tilde{\chi}^-_i\tilde{\chi}^0_j$ 
($i=1,2$ and $j=1$ to $4$) for the sake of generality. \\

The parton--level production process $d\bar{u}\rightarrow\tilde{\chi}^-_i
\tilde{\chi}^0_j$ is generated by the three mechanisms shown in 
Fig.~\ref{fig:fig3}: the $s$--channel $W^-$ exchange, the $t$--channel 
$\tilde{d}_L$ exchange and the $u$--channel $\tilde{u}_L$ exchange.  
Before presenting the explicit form for the production helicity amplitudes,
we note that the chirality mixing of the first and second generation sfermions 
is proportional to the fermion mass much smaller than the beam energy of the
Tevatron.  Therefore, we can safely ignore the sfermion left--right 
chirality mixing in calculating the associated chargino-neutralino production 
rate so that the trilinear term $A_f$ does not play any role in  
the high energy process.  
The phase $\Phi_3$ of the gluino mass is irrelevant as well. 
Consequently, this associated production of chargino and neutralino
and their subsequent decays involve only two CP--violating
phases $\{ \Phi_\mu, \Phi_1 \}$.
With these good approximations and after an appropriate Fierz
transformation to the $\tilde{u}_L$ and $\tilde{d}_L$--exchange amplitudes
the transition matrix element can be written in the form  
\begin{eqnarray}
T\left(d\bar{u}\rightarrow\tilde{\chi}^-_i\tilde{\chi}^0_j\right)
 = \frac{e^2}{s}Q^{ij}_{\alpha\beta}
   \left[\bar{v}(\bar{u})  \gamma_\mu P_\alpha  u(d)\right]
   \left[\bar{u}(\tilde{\chi}^-_i) \gamma^\mu P_\beta 
               v(\tilde{\chi}^0_j)\right]\,.
\label{eq:production amplitude}
\end{eqnarray}
Here $Q_{\alpha\beta}$ are the so--called generalized bilinear 
charges \cite{SZ}, 
classified according to the chiralities $\alpha,\beta=L,R$ of the associated 
quark current and the chargino/neutralino current. The explicit forms of 
these bilinear charges are  
\begin{eqnarray}
&& Q^{ij}_{LL}= \frac{D_W}{\sqrt{2}s_W^2}{\cal W}_{Lij}
              -\frac{D^{\tilde{d}_L}_{u}}{\sqrt{2}s_W}g_{Lij},\nonumber\\ 
&& Q^{ij}_{LR}= \frac{D_W}{\sqrt{2}s_W^2}{\cal W}_{Rij}
              +\frac{D^{\tilde{u}_L}_{t}}{\sqrt{2}s_W}g_{Rij},\nonumber\\[1mm]
&& Q^{ij}_{RL}= 0,\qquad
   Q^{ij}_{RR}= 0 .
\label{eq:bilinear charge for production}
\end{eqnarray}
with the $s$--, $t$--, and $u$--channel propagators:
\begin{eqnarray}
D_W=\frac{s}{s-m^2_W+im_W\Gamma_W}\,, \qquad
D^{\tilde{u}_L}_t=\frac{s}{t-m^2_{\tilde{u}_L}}\,, \qquad \
D^{\tilde{d}_L}_u=\frac{s}{u-m^2_{\tilde{d}_L}}\,,
\end{eqnarray}
where $s=(p_d+p_{\bar{u}})^2$, $t=(p_d-p_{\tilde{\chi}^-_i})^2$ and
$u=(p_d-p_{\tilde{\chi}^0_j})^2$, and the couplings ${\cal W}_{Lij}$,
${\cal W}_{Rij}$, $g_{Lij}$ and $g_{Rij}$ are given by 
\begin{eqnarray}
&& {\cal W}_{Lij}=U_{Li1} N^*_{j2} 
              + \frac{1}{\sqrt{2}}U_{Li2} N^*_{j3} ,\nonumber\\
&& {\cal W}_{Rij}=U_{Ri1} N_{j2} 
              - \frac{1}{\sqrt{2}}U_{Ri2} N_{j4}   ,\nonumber\\
&& g_{Lij}=\frac{U_{Li1}}{s_W c_W}
   \left[s_W (Q_d+\frac{1}{2})N^*_{j1}-\frac{1}{2}c_W N^*_{j2}
   \right] ,\nonumber\\
&& g_{Rij}=\frac{U_{Ri1}}{s_W c_W}
   \left[s_W (Q_u-\frac{1}{2})N_{j1}+\frac{1}{2}c_W N_{j2} 
   \right].
\label{eq:production bilinear charges}
\end{eqnarray}
Here $U_{Lij,Rij}$ ($N_{ij}$) are the chargino (neutralino)
mixing matrix elements and $Q_d=-1/3$ and $Q_u=+2/3$ the electric charges 
for the down-- and up--type quarks, respectively. Note that the coefficients 
$g_{Lij}$ and $g_{Rij}$ for the $t$-- and $u$--channel diagrams are governed 
only by gaugino components of the chargino and neutralino while 
${\cal W}_{Lij}$ and ${\cal W}_{Rij}$ are determined by both the gaugino 
and higgsino components.\\

Defining the $\tilde{\chi}^-_i$ flight direction with respect to the
down quark momentum direction by $\Theta$, the explicit form of
the production helicity amplitudes 
can be determined from eq.~(\ref{eq:production amplitude}). 
In the limit of neglecting the initial $d$ and $u$ quark masses,  
the $d$ and $\bar{u}$ helicities are opposite 
to each other in all the exchange amplitudes, but the $\tilde{\chi}^-_i$ 
and $\tilde{\chi}^0_j$ helicities are less correlated due to the non--zero 
masses of the particles; the amplitudes with equal chargino/neutralino 
helicities $\propto m_{\tilde{\chi}^-_i,\tilde{\chi}^0_j}
/\sqrt{s}$ must vanish only for asymptotic energies. Denoting 
the down quark helicity by the first index, the $\tilde{\chi}^-_i$ and 
$\tilde{\chi}^0_j$ helicities 
by the remaining two indices, the production helicity amplitudes 
$T(\sigma;\lambda_i,\lambda_j)=2\pi\alpha\,\langle\sigma;\lambda_i\lambda_j
\rangle$ can be derived by the so--called 2--component spinor
technique \cite{HZ}, which yields 
\begin{eqnarray}
&& \langle +;++\rangle 
   =-\left[Q^{ij}_{RR}\sqrt{1-\eta^2_+}+Q^{ij}_{RL}\sqrt{1-\eta^2_-}\right]
     \sin\Theta\,,\nonumber\\
&& \langle +;+-\rangle 
   =-\left[Q^{ij}_{RR}\sqrt{(1+\eta_+)(1+\eta_-)}+
           Q^{ij}_{RL}\sqrt{(1-\eta_+)(1-\eta_-)}\right](1+\cos\Theta)
               \,,\nonumber\\
&& \langle +;-+\rangle 
   =+\left[Q^{ij}_{RR}\sqrt{(1-\eta_+)(1-\eta_-)}+
           Q^{ij}_{RL}\sqrt{(1+\eta_+)(1+\eta_-)}\right](1-\cos\Theta)
              \,, \nonumber\\
&& \langle +;--\rangle 
   =+\left[Q^{ij}_{RR}\sqrt{1-\eta^2_-}+Q^{ij}_{RL}\sqrt{1-\eta^2_+}\right]
     \sin\Theta\,,\nonumber\\ 
&& \langle -;++\rangle 
   =-\left[Q^{ij}_{LR}\sqrt{1-\eta^2_+}+Q^{ij}_{LL}\sqrt{1-\eta^2_-}\right]
     \sin\Theta \,,\nonumber\\
&& \langle -;+-\rangle 
   =+\left[Q^{ij}_{LR}\sqrt{(1+\eta_+)(1+\eta_-)}+
           Q^{ij}_{LL}\sqrt{(1-\eta_+)(1-\eta_-)}\right](1-\cos\Theta)
               \,,\nonumber\\
&& \langle -;-+\rangle 
   =-\left[Q^{ij}_{LR}\sqrt{(1-\eta_+)(1-\eta_-)}+
           Q^{ij}_{LL}\sqrt{(1+\eta_+)(1+\eta_-)}\right](1+\cos\Theta)
              \,, \nonumber\\
&& \langle -;--\rangle 
   =+\left[Q^{ij}_{LR}\sqrt{1-\eta^2_-}+Q^{ij}_{LL}\sqrt{1-\eta^2_+}\right]
     \sin\Theta\,, 
\label{eq:helicity amplitude}
\end{eqnarray}
where $\eta_\pm=\lambda^{1/2}(1,\mu^2_i,\mu^2_j)\pm(\mu^2_i-\mu^2_j)$
with $\mu^2_i\,(\mu^2_j)=m^2_{\tilde{\chi}^-_i}(m^2_{\tilde{\chi}^0_j}/s)$ 
and $\lambda(x,y,z)=x^2+y^2+z^2-2xy-2yz-2zx$. 
If the arguments are not specified, then the notation $\lambda$ stands 
for $\lambda(1,\mu^2_i,\mu^2_j)$ in the following. \\

All physical observables constructed through the production process are 
determined by the production helicity amplitudes and expressed 
in a very simple form by sixteen quartic charges \cite{SZ} 
containing the full information on the dynamical properties of 
the production process. These quartic charges are expressed in terms 
of the bilinear 
charges $Q^{ij}_{\alpha\beta}$ given by eq.~(\ref{eq:bilinear charge 
for production}) and classified according to their transformation 
properties under parity (P) as follows:\\
\vskip 0.2cm
\noindent
{\bf (a)} \underline{\bf Eight P--even terms}:
\begin{eqnarray}
&& Q^{ij}_1=\frac{1}{4}
            \left[|Q^{ij}_{RR}|^2+|Q^{ij}_{LL}|^2
                 +|Q^{ij}_{RL}|^2+|Q^{ij}_{LR}|^2\right]\,,\nonumber\\
&& Q^{ij}_2=\frac{1}{2}\,\real\!
            \left[Q^{ij}_{RR}Q^{ij*}_{RL}
                 +Q^{ij}_{LL}Q^{ij*}_{LR}\right]\,,\nonumber\\
&& Q^{ij}_3=\frac{1}{4}
            \left[|Q^{ij}_{LL}|^2+|Q^{ij}_{RR}|^2
                 -|Q^{ij}_{RL}|^2-|Q^{ij}_{LR}|^2\right]\,,\nonumber\\
&& Q^{ij}_4=\frac{1}{2}\,\imag\!
            \left[Q^{ij}_{RR}Q^{ij*}_{RL}
                 +Q^{ij}_{LL}Q^{ij*}_{LR}\right]\,,\nonumber\\
&& Q^{ij}_5=\frac{1}{2}\,\real\!
            \left[Q^{ij}_{RR}Q^{ij*}_{LR}
                 +Q^{ij}_{LL}Q^{ij*}_{RL}\right]\,,\nonumber\\
&& Q^{ij}_6=\frac{1}{2}\,\imag\!
            \left[Q^{ij}_{RR}Q^{ij*}_{LR}
                 +Q^{ij}_{LL}Q^{ij*}_{RL}\right]\,,\nonumber\\
&& Q^{ij}_7=\real\!\left[Q^{ij}_{RR}Q^{ij*}_{LL}\right]\,,\nonumber\\
&& Q^{ij}_8=\real\!\left[Q^{ij}_{RL}Q^{ij*}_{LR}\right]\,,
\end{eqnarray}\\
{\bf (b)} \underline{\bf Eight P--odd terms}: 
\begin{eqnarray}
&& Q^{'ij}_1=\frac{1}{4}
             \left[|Q^{ij}_{RR}|^2+|Q^{ij}_{RL}|^2
                  -|Q^{ij}_{LR}|^2-|Q^{ij}_{LL}|^2\right]\,,\nonumber\\
&& Q^{'ij}_2=\frac{1}{2}\,\real\!
             \left[Q^{ij}_{RR}Q^{ij*}_{RL}
                  -Q^{ij}_{LL}Q^{ij*}_{LR}\right]\,,\nonumber\\
&& Q^{'ij}_3=\frac{1}{4}
             \left[|Q^{ij}_{RR}|^2+|Q^{ij}_{LR}|^2
                  -|Q^{ij}_{RL}|^2-|Q^{ij}_{LL}|^2\right]\,,\nonumber\\
&& Q^{'ij}_4=\frac{1}{2}\,\imag\!
             \left[Q^{ij}_{RR}Q^{ij*}_{RL}
                  -Q^{ij}_{LL}Q^{ij*}_{LR}\right]\,,\nonumber\\
&& Q^{'ij}_5=\frac{1}{2}\,\real\!
             \left[Q^{ij}_{RR}Q^{ij*}_{LR}
                  -Q^{ij}_{LL}Q^{ij*}_{RL}\right]\,,\nonumber\\
&& Q^{'ij}_6=\frac{1}{2}\,\imag\!
             \left[Q^{ij}_{RR}Q^{ij*}_{LR}
                  -Q^{ij}_{LL}Q^{ij*}_{RL}\right]\,,\nonumber\\
&& Q^{'ij}_7=\imag\!\left[Q^{ij}_{RR}Q^{ij*}_{LL}\right]\,,\nonumber\\
&& Q^{'ij}_8=\imag\!\left[Q^{ij}_{RL}Q^{ij*}_{LR}\right]\,.
\end{eqnarray}
We note that these 16 quartic charges comprise the most complete set for
any fermion--pair production process in $d\bar{u}$ collisions when the quark 
masses are neglected. On the other hand, the quartic charges defined by an 
imaginary part of the bilinear--charge correlations might be {\it
non--vanishing} only when there exist complex CP--violating couplings 
or/and CP--preserving phases like rescattering phases or finite widths 
of the intermediate particles. Therefore, such nonvanishing 
values of those quartic charges may signal CP violation in a given process.

\subsection{Parton--level production cross section}
\label{subsec:4.3}

The unpolarized differential production cross section of the parton--level 
process $d\bar{u}\rightarrow\tilde{\chi}^-_i\tilde{\chi}^0_j$ 
is obtained straightforwardly by taking the average/sum 
over the initial/final helicities:
\begin{eqnarray}
\frac{{\rm d}\hat{\sigma}}{{\rm d}\cos\Theta}
      (d\bar{u}\rightarrow\tilde{\chi}^-_i\tilde{\chi}^0_j)
 =\frac{\pi\alpha^2}{32 s} \lambda^{1/2} \, 
  \sum_{\sigma\lambda_i\lambda_j} |\langle\sigma;\lambda_i\lambda_j
\rangle|^2\,.
\end{eqnarray}
Carrying out the sum, one finds the following expression for the parton level 
differential cross section in terms of the quartic charges:
\begin{eqnarray}
\frac{{\rm d}\hat{\sigma}}{{\rm d}\cos\Theta}
      (d\bar{u}\rightarrow\tilde{\chi}^-_i\tilde{\chi}^0_j)
&=&\frac{\pi\alpha^2}{8 s} \lambda^{1/2} 
  \bigg\{[4-(\eta_+-\eta_-)^2+(\eta_++\eta_-)^2\cos^2\Theta]\, Q^{ij}_1\,,
       \nonumber\\
&&+4\sqrt{(1-\eta^2_+)(1-\eta^2_-)}\, Q^{ij}_2
             +4(\eta_++\eta_-)\cos\Theta\, Q^{ij}_3\bigg\}\,.
\label{eq:cross section}
\end{eqnarray}
Thus, the three P--even quartic charges $Q^{ij}_1$, $Q^{ij}_2$ and $Q^{ij}_3$ 
determine the $\Theta$ dependence of the cross section completely.\\

The final cross section of the $\tilde{\chi}^-_i\tilde{\chi}^0_j$
production in $p\bar{p}$ collisions is obtained by convoluting the parton level 
cross section with a  
parton distribution and it is given by, 
\begin{eqnarray}
\sigma (p\bar p \to \tilde{\chi}^-_1 \tilde\chi^0_2+X) 
   = \frac{\kappa}{3}\int_0^1\frac{dx}{x}\int_{\tau_{\rm min}}^1 d\tau 
     \left[f_{d/p}(x) f_{\bar{u}/\bar{p}}(\tau/x)
          +f_{\bar{u}/p}(x) f_{d/\bar{p}}(\tau/x)\right]
     \hat{\sigma}(\hat{s}=\tau s)\,, 
\label{eq:inclusive x-section}
\end{eqnarray}
where $f$ are the respective parton fluxes in $p$ and $\bar p$ and
$\tau_{\rm min}=(m_{\tilde{\chi}^-_1}+m_{\tilde{\chi}^0_2})^2/s$. We have used 
the CTEQ4m \cite{CTEQ4m} parametrisation to obtain parton distribution
setting the relevant QCD scale to the c.m. energy $\hat s$ of the parton 
level process.  We take into account the dominant QCD radiative corrections 
to the production cross section by taking the parameter $\kappa=1.3$ 
\cite{SPIRA} in eq.~(\ref{eq:inclusive x-section}). The production cross 
section for the associated positively--charged  
chargino and neutralino pair is the same as its charge--conjugate one. \\  

Figure~\ref{fig:fig6} shows the distribution of the production cross section 
$\sigma(p\bar{p}\rightarrow
\tilde{\chi}^-_1\tilde{\chi}^0_2+X)$ with the scattering angle $\Theta$ 
for various sets of $\{\Phi_\mu,\Phi_1\}$ in the scenario ${\cal S}1$ and 
${\cal S}2$ for a given c.m. energy of 1.8 TeV. We find that the production 
cross section is very sensitive to $\Phi_\mu$, but (almost) insensitive to 
$\Phi_1$. In order to look into this feature more clearly, we present in
Fig.~\ref{fig:fig7} the integrated production cross section 
$\sigma(p\bar{p}\rightarrow
\tilde{\chi}^-_1\tilde{\chi}^0_2+X)$ on the $\{\Phi_\mu,\Phi_1\}$ plane 
in (a) ${\cal S}1$ and (b) ${\cal S}2$, for which the constraints from
the electron EDM are embedded on the plane.
The results exhibit a few interesting aspects:
\begin{itemize}

\item The differential cross section is (almost) forward-backward symmetric
      in both scenarios. This can be understood by noting the fact that the 
      higgsino parts of ${\cal W}_{L,R}$ are suppressed in both scenarios 
      due to large values of $|\mu|$ and  
      that the relevant quartic charges $Q_{1,2}$ 
      are forward--backward symmetric but $Q_3$ is proportional 
      to $\cos\Theta$ to a good approximation due to the assumption
      $m_{\tilde u} = m_{\tilde d}$.

\item The large $|\mu|$ and small squark masses in the scenario ${\cal S}2$ 
      reduce the production cross section due to a 
      severe destructive interference between the $W$--exchange and  
      squark--exchange diagrams.
      On the other hand, in the scenario ${\cal S}1$,  
      the $t$-- and $u$--channel contributions with the large squark masses
      can be ignored so that  
      the lack of destructive interference yields a much larger value
      of the production cross section in the scenario.

\item Except for the region of $\Phi_\mu=0,2\pi$ in ${\cal S}1$, the
      integrated production cross section is almost independent of the phase 
      $\Phi_1$ and it decreases as $\Phi_\mu$ approaches $\pi$ in 
      both scenarios. This property is mainly set by the dependence of
      the chargino and neutralino masses on the CP--violating phases.

\end{itemize}
Consequently, the production cross section itself can be a very sensitive 
probe to the phase $\Phi_\mu$, but not to the phase $\Phi_1$.

\subsection{Chargino and neutralino polarization vectors}
\label{subsec:4.4}

The spin--1/2 chargino and neutralino through the parton--level
process $d\bar{u} \rightarrow\tilde{\chi}^-_i\tilde{\chi}^0_j$ are 
in general polarized and their polarizations are reflected in the 
distributions of their decay products; 
$\tilde{\chi}^-_i\rightarrow \tilde\chi^0_1\,\ell^-\bar{\nu}_\ell$ and 
$\tilde{\chi}^0_j\rightarrow \tilde{\chi}^0_1\,
\ell^{\prime +}\ell^{\prime -}$.  
In order to have a rough estimate of the
effects of the CP--violating phases on the spin correlations,
it is meaningful to investigate the chargino and neutralino 
polarizations in the parton--level production process 
$d\bar{u} \rightarrow\tilde{\chi}^-_i
\tilde{\chi}^0_j$. \\

The polarization vector $\overrightarrow{\cal P}^{\tilde{\chi}^-_i}
=({\cal P}^{\tilde{\chi}^-_i}_L,{\cal P}^{\tilde{\chi}^-_i}_T,
{\cal P}^{\tilde{\chi}^-_i}_N)$ of the produced chargino 
$\tilde{\chi}^-_i$ is defined 
in the rest frame in which the axis $\hat{z}\| L$ is in the flight direction 
of $\tilde{\chi}^-_i$, $\hat{x}\| T$ rotated counter--clockwise 
in the production plane, and $\hat{y}=\hat{z}\times\hat{x}\| N$
of the decaying chargino $\tilde{\chi}^-_i$. 
Accordingly, ${\cal P}^{\tilde{\chi}^-_i}_L$ denotes the 
component parallel to the $\tilde{\chi}^-_i$ flight direction in the c.m. 
frame, ${\cal P}^{\tilde{\chi}^-_i}_T$ the transverse component 
in the production plane, and 
${\cal P}^{\tilde{\chi}^-_i}_N$ the component normal to the production plane. 
The three components of the chargino polarization vector can be expressed by 
the production helicity amplitudes (\ref{eq:helicity amplitude}) as
\begin{eqnarray}
&& {\cal P}^{\tilde{\chi}^-_i}_L=\frac{1}{4}\sum_{\sigma=\pm}\left\{
              |\langle\sigma;++\rangle|^2+|\langle\sigma;+-\rangle|^2
             -|\langle\sigma;-+\rangle|^2-|\langle\sigma;--\rangle|^2
                                           \right\}/{\cal N}
              \,,\nonumber\\
&& {\cal P}^{\tilde{\chi}^-_i}_T=\frac{1}{2}\real
              \bigg\{\sum_{\sigma=\pm}\left[
              \langle\sigma;++\rangle\langle\sigma;-+\rangle^*
             +\langle\sigma;--\rangle\langle\sigma;+-\rangle^*
                          \right]\bigg\}/{\cal N}\,,\nonumber\\
&& {\cal P}^{\tilde{\chi}^-_i}_N=\frac{1}{2}\imag
              \bigg\{\sum_{\sigma=\pm}\left[
              \langle\sigma;--\rangle\langle\sigma;+-\rangle^*
             -\langle\sigma;++\rangle\langle\sigma;-+\rangle^*
                           \right]\bigg\}/{\cal N}\,,
\label{eq:cpolvec}
\end{eqnarray}
with the normalization corresponding to the unpolarized distribution
\begin{eqnarray}
{\cal N}=\frac{1}{4}\sum_{\lambda_i\lambda_j}
           \left[|\langle +;\lambda_i\lambda_j\rangle|^2
                +|\langle -;\lambda_i\lambda_j\rangle|^2\right]\,.
\end{eqnarray}
The polarization vector of the neutralino $\tilde{\chi}^0_j$
can be obtained similarly from the production helicity amplitudes
(\ref{eq:helicity amplitude})
by exchanging the chargino and neutralino helicities in eq.~(\ref{eq:cpolvec})
\begin{eqnarray}
&& {\cal P}^{\tilde{\chi}^0_j}_L=\frac{1}{4}\sum_{\sigma=\pm}\left\{
              |\langle\sigma;++\rangle|^2+|\langle\sigma;-+\rangle|^2
             -|\langle\sigma;+-\rangle|^2-|\langle\sigma;--\rangle|^2
                                           \right\}/{\cal N}
              \,,\nonumber\\
&& {\cal P}^{\tilde{\chi}^0_j}_T=\frac{1}{2}\real
              \bigg\{\sum_{\sigma=\pm}\left[
              \langle\sigma;++\rangle\langle\sigma;+-\rangle^*
             +\langle\sigma;--\rangle\langle\sigma;-+\rangle^*
                          \right]\bigg\}/{\cal N}\,,\nonumber\\
&& {\cal P}^{\tilde{\chi}^0_j}_N=\frac{1}{2}\imag
              \bigg\{\sum_{\sigma=\pm}\left[
              \langle\sigma;--\rangle\langle\sigma;-+\rangle^*
             -\langle\sigma;++\rangle\langle\sigma;+-\rangle^*
                           \right]\bigg\}/{\cal N}\,.
\end{eqnarray}\\[-4mm]

The longitudinal and transverse components of the polarization 
vectors are P--odd and CP--even, but  
the normal component is P--even and CP--odd so that  the normal 
polarization component can be generated only by the complex
production amplitudes. Certainly there exist non--trivial phases 
in CP non--invariant SUSY models. 
Also the non--zero width of the $Z$ boson and loop corrections generate 
non--trivial phases; however, the $Z$ boson width contribution to the
normal polarization is negligible for high energies as
mentioned before, and so are the radiative corrections. 
As a result the normal component is effectively generated by the genuine
CP--odd phases of the couplings.\\ 

It is straightforward to find the normal polarization components of
$\tilde{\chi}^-_i$ and $\tilde{\chi}^0_j$: 
\begin{eqnarray}
{\cal P}^{\tilde{\chi}^-_i}_N 
          = \frac{8}{\cal N}\,\lambda^{1/2}\mu_i\sin\Theta\, Q_4^{ij}\,,
            \qquad
{\cal P}^{\tilde{\chi}^0_j}_N
          = \frac{8}{\cal N}\,\lambda^{1/2}\mu_j\sin\Theta\, Q_4^{ij}\,.
\label{eq:polarization vector}
\end{eqnarray}
As expected from the CP properties of the normal components,
they are determined by the CP--odd quartic charge $Q_4^{ij}$,
which requires some complex couplings. In order to estimate the size
of the normal polarization quantitatively, we present in Fig.~\ref{fig:fig8}
the normal polarization ${\cal P}^{\tilde{\chi}^0_2}_N$ for four
different combinations of $\{\Phi_\mu,\Phi_1\}$ in the scenarios 
${\cal S}1$ and ${\cal S}2$ taking the parton--level c.m. energy of 300 GeV
for the sake of illustration.
[The chargino normal polarization is proportional 
to the neutralino normal polarization.] The expression 
(\ref{eq:polarization vector}) 
and Fig.~\ref{fig:fig8} lead us to the following features: 
\begin{itemize}
\item The normal polarizations are suppressed near the thresholds or at
      high energies.
\item In the scenario ${\cal S}1$ with large squark masses, the quartic
      charge $Q_4^{ij}$ is forward-backward symmetric so that the
      normal polarizations also are forward--backward symmetric.
      However, this property is not maintained in the scenario
      ${\cal S}2$.
\item In both scenarios, the size of the normal polarizations is 
      too small to measure the CP--violating 
      phases directly in the production of the associated chargino
      and neutralino pair.
\end{itemize}
To conclude, it is likely that after applying the stringent EDM constraints 
to the CP--violating phases, we could not expect to have the normal 
polarizations of the chargino and neutralino large enough to be measured 
at the Tevatron.

\section{Polarized Chargino and Neutralino Decays}
\label{sec: chargino and neutralino decay}

The detection efficiency of the tri--lepton signatures at the Tevatron relies 
crucially on the branching fractions and distributions of 
the final three leptons, the latter of which depend on the
polarizations of the decaying chargino and neutralino. 
To estimate the branching fractions, we need to calculate all
the main partial decay widths that depend on the sparticle
and Higgs--boson spectra in the MSSM. This section is devoted to
a comprehensive discussion on the chargino and neutralino leptonic
decays, including the polarizations of the decaying chargino and 
neutralino and the branching fractions. Firstly, we present
the chargino and neutralino decay amplitudes in terms of the corresponding
bilinear charges and the polarized decay distributions in terms
of the quartic charges. Secondly, we calculate the decay density matrices
by using the so--called Bouchiat--Michel formulas \cite{BM}, that are 
to be used to form the complete production--decay spin/angular
correlations. Finally, we estimate the branching fractions of the
leptonic decays $\tilde{\chi}^-_1\rightarrow\tilde{\chi}^0_1\,\ell^-
\bar{\nu}_\ell$ and $\tilde{\chi}^0_2\rightarrow\tilde{\chi}^0_1
\,\ell^{\prime +}\ell^{\prime -}$
in the scenarios ${\cal S}1$ and ${\cal S}2$.

\subsection{Polarized decay distributions}
\label{subsec:5.1}

The diagrams contributing to the process 
$\tilde{\chi}^-_i\rightarrow \tilde{\chi}^0_1\,\ell^-\bar{\nu}_\ell$ and 
$\tilde{\chi}^0_j\rightarrow \tilde{\chi}^0_1\,\ell^{\prime -}\ell^{\prime +}$ 
with $\ell,\ell^{\prime}=e$, $\mu$ are shown in Figs.~\ref{fig:fig9}(a) 
and (b).  Here, the exchange of the neutral and charged Higgs bosons 
[replacing the $W^-$ and $Z$ bosons] are neglected since the Yukawa 
couplings to the light first and second generation leptons are very small. 
In this case, all the components of the decay matrix elements 
are, after a simple Fierz transformation, written as
\begin{eqnarray}
{\cal D}\left(\tilde{\chi}^-_i\rightarrow\tilde{\chi}^0_1\,
              \ell^-\bar{\nu}_\ell\right)
 &=& \frac{e^2}{s'}\,C^i_{\alpha\beta}
   \left[\bar{u}(\tilde{\chi}^0_1) \gamma^\mu P_\alpha 
               u(\tilde{\chi}^-_i)\right]
   \left[\bar{u}(\ell^-)  \gamma_\mu P_\beta  v(\bar{\nu}_\ell)\right]\,,
   \nonumber \\
{\cal D}\left(\tilde{\chi}^0_j\rightarrow\tilde{\chi}^0_1\, \ell^{\prime -}
       \ell^{\prime +}\right)
 &=& \frac{e^2}{s''}\, N^j_{\alpha\beta}
   \left[\bar{v}(\tilde{\chi}^0_j) \gamma^\mu P_\alpha
               v(\tilde{\chi}^0_1)\right]
   \left[\bar{u}(\ell^{\prime -})  \gamma_\mu P_\beta  
               v(\ell^{\prime +})\right]\,,
\label{eq:decay amplitude}
\end{eqnarray}
where $\alpha,\beta=L,R$.
Note that since the decaying neutralino is treated as an anti-particle
in the associated production process $d\bar{u}\rightarrow\tilde{\chi}^-_i
\tilde{\chi}^0_j$, the $v$ spinors for the decaying neutralino and
the LSP appear in the expression for the neutralino leptonic decay.
However, owing to the Majorana property of the neutralinos it
does not matter whether the decay neutralino is treated as a particle
or an anti-particle. The generalized bilinear charges 
$C^i_{\alpha\beta}$ for the chargino leptonic decay are given by
\begin{eqnarray}
&& C^i_{LL}=+\frac{D_W'}{\sqrt{2}s_W^2}{\cal W}_{Li1}^*
              -\frac{D^{\tilde{l}_L}_{u'}}{\sqrt{2}s_W}g^*_{Li1},\nonumber\\
&& C^i_{RL}=-\frac{D_W'}{\sqrt{2}s_W^2}{\cal W}_{i1R}^*
              +\frac{D^{\tilde{\nu}}_{t'}}{\sqrt{2}s_W}g^*_{Ri1},\nonumber\\
&& C^i_{LR}= C^i_{RR} = 0,
\label{eq:cbil}
\end{eqnarray}
The couplings ${\cal W}_{Li1}$, ${\cal W}_{Ri1}$, $g_{Li1}$, and
$g_{Ri1}$ are given in eq.~(\ref{eq:production bilinear charges}). 
The $s'$--, $t'$-- and $u'$--channel propagators are
\begin{eqnarray}
D'_W=\frac{s'}{s'-m^2_W+im_W\Gamma_W}\,,\qquad
D^{\tilde{\nu}}_{t'}=\frac{s'}{t'-m^2_{\tilde{\nu}}}\,,\qquad
D^{\tilde{\l}_L}_{u'}=\frac{s'}{u'-m^2_{\tilde{l}_L}}\,.
\end{eqnarray}
where the Mandelstam variables $s'$, $t'$ and $u'$ are
defined in terms of the 4-momenta, $q_0,\, q$ and $\bar{q}$,
of $\tilde{\chi}^0_1$, $\ell^-$ and $\bar{\nu}_\ell$, respectively,
as
\begin{eqnarray}
s'=(q+\bar{q})^2 \ \ , \ \ t'=(q_0+\bar{q})^2 \ \ , 
\ \ u'=(q_0 +q)^2\,,
\end{eqnarray}
On the other hand, the generalized bilinear charges 
$N^j_{\alpha\beta}$ for the leptonic decay $\tilde\chi_j^0 \rightarrow
\tilde{\chi}^0_1\,\ell^{\prime +} \ell^{\prime -}$ are given by
\begin{eqnarray}
&& N^j_{LL}=+\frac{D_Z''}{s_W^2c_W^2}(s_W^2 -\frac{1}{2}){\cal Z}_{j1}
              -D^{\tilde{\ell}^{\prime}_L}_{t''}\,h_{Lj1}\,,\nonumber\\ 
&& N^j_{LR}=+\frac{D_Z''}{c_W^2}{\cal Z}_{j1}
              +D^{\tilde{\ell}^{\prime}_R}_{u''}\,h_{Rj1}\,,\nonumber\\
&& N^j_{RL}=-\frac{D_Z''}{s_W^2c_W^2}(s_W^2 -\frac{1}{2}){\cal Z}^*_{j1}
              +D^{\tilde{\ell}^{\prime}_L}_{u''}\,h^*_{Lj1}\,,\nonumber\\
&& N^j_{RR}=-\frac{D_Z''}{c_W^2}{\cal Z}^*_{j1}
              -D^{\tilde{\ell}^{\prime}_R}_{t''}\,h^*_{Rj1}\,,
\end{eqnarray}
where the $s''$--, $t''$-- and $u''$--channel propagators are 
\begin{eqnarray}
D_Z''=\frac{s''}{s''-m^2_Z+im_Z\Gamma_Z},\qquad
D^{\tilde{\l}_{L,R}}_{t''}=\frac{s''}{t''-m^2_{\tilde{l}_{L,R}}},\qquad
D^{\tilde{\l}_{L,R}}_{u''}=\frac{s''}{u''-m^2_{\tilde{l}_{L,R}}},
\end{eqnarray}
with the Mandelstam variables $s''=(q'+\bar{q}')^2$, $t''=(q_0'+\bar{q}')^2$ 
and $u''=(q_0'+q')^2$ in terms of the 4--momenta,
$q^\prime_0,\, q^\prime$ and $\bar{q}^\prime$, of $\tilde{\chi}^0_1$, 
$\ell^{\prime -}$ and $\ell^{\prime +}$, respectively. The 
couplings ${\cal Z}_{ij}$,
$h_{Lij}$ and $h_{Rij}$ are expressed in terms of neutralino 
diagonalization matrix elements
$N_{ij}$ as
\begin{eqnarray}
&& {\cal Z}_{ij}=\frac{1}{2}
                 \left[N_{i3}N^*_{j3}-N_{i4}N^*_{j4}\right]\,,\nonumber\\
&& h_{Lij}=\frac{1}{4 s_W^2c_W^2}(N_{i2}c_W+N_{i1}s_W)(N^*_{j2}c_W+N^*_{j1}s_W)
                 \,,\nonumber\\
&& h_{Rij}=\frac{1}{c_W^2}N_{i1}N^*_{j1}\,,
\end{eqnarray}
and they satisfy the Hermiticity relations reflecting the CP relations
\begin{eqnarray}
{\cal Z}_{ij}={\cal Z}^*_{ji}\,,\qquad
h_{Lij}=h^*_{Lji}\,,\qquad
h_{Rij}=h^*_{Rji}\,.
\end{eqnarray}
Note that ${\cal Z}_{j1}$ is governed by the higgsino components of
$\tilde{\chi}^0_j$ while $h_{Lj1}$ and $h_{Rj1}$ are determined by 
the gaugino components of the neutralino. Therefore, as will be shown
later, ${\cal Z}_{21}$ are suppressed in the scenario ${\cal S}2$ with
a large $|\mu|$, while the $t$-- and $u$--channel diagrams are
suppressed in the scenario ${\cal S}1$ with large selectron
masses.\\

Applying the polarization projection operator of the chargino 
to the amplitude squared of the chargino leptonic decay 
$\tilde{\chi}^-_i\rightarrow \tilde{\chi}^0_1\,\ell^-\bar{\nu}_\ell$
yields the polarized decay distributions with the
polarization vector $n_i^\mu$ of the chargino $\tilde{\chi}^-_i$:  
\begin{eqnarray}
|{\cal D}|^2(n_i) &=& 
-4(t'- m_{\chi_i^-}^2)(t'- m_{\chi_1^0}^2)(C^i_1-C^i_3) 
-4(u'- m_{\chi_i^-}^2)(u'- m_{\chi_1^0}^2)(C^i_1+C^i_3)   \nonumber\\
&& -8 m_{\tilde{\chi}^-_i} m_{\tilde{\chi}^0_1} s'C^i_2 \nonumber \\
&& -8 (n_i\cdot\bar{q})\left[
 m_{\tilde{\chi}^-_i}(m_{\tilde{\chi}^0_1}^2-u')(C^{i\prime}_1+C^{i\prime}_3)
+m_{\tilde{\chi}^0_1}(m_{\tilde{\chi}^-_i}^2-t')C^{i\prime}_2\right]\nonumber\\
&& +8(n_i\cdot q)\left[
-m_{\tilde{\chi}^-_i}(m_{\tilde{\chi}^0_1}^2-t')(C^{i\prime}_1-C^{i\prime}_3)
+m_{\tilde{\chi}^0_1}(m_{\tilde{\chi}^-_i}^2-u')C^{i\prime}_2\right]\nonumber\\ 
&& -16\, m_{\tilde{\chi}^0_1}\langle q_i n_i q \bar{q}\rangle C^i_4\,,
\label{eq:decay distribution}
\end{eqnarray}
where $\langle q_i n_i q_1 \bar{q}_2\rangle\equiv\epsilon_{\mu\nu\rho\sigma} 
q_i^{\mu}n_i^{\nu}q^{\rho}\bar{q}^{\sigma}$ with the convention 
$\epsilon_{0123}=+1$.
Here, the quartic charges $\{C_{i1}$ - $C_{i4}\}$ and 
$\{C'_{i1}$ - $C'_{i3}\}$ for the chargino decays are defined by
\begin{eqnarray}
&& C^i_{1}=\frac{1}{4}
            \left[|C^i_{RR}|^2+|C^i_{LL}|^2
                 +|C^i_{RL}|^2+|C^i_{LR}|^2\right]\,,  \nonumber\\
&& C^i_{2}=\frac{1}{2}\real
            \left[C^i_{RR}C^{i*}_{LR}
                 +C^i_{LL}C^{i*}_{RL}\right]\,,       \nonumber\\
&& C^i_{3}=\frac{1}{4}
            \left[|C^i_{LL}|^2+|C^i_{RR}|^2
                 -|C^i_{RL}|^2-|C^i_{LR}|^2\right]\,,  \nonumber\\
&& C^i_{4}=\frac{1}{2}\imag
            \left[C^i_{RR}C^{i*}_{LR}
                 +C^i_{LL}C^{i*}_{RL}\right]\,,       \nonumber\\
&& C^{i\prime}_{1}=\frac{1}{4}
             \left[|C^i_{RR}|^2+|C^i_{RL}|^2
                  -|C^i_{LR}|^2-|C^i_{LL}|^2\right]\,, \nonumber\\
&& C^{i\prime}_{2}=\frac{1}{2}\real
             \left[C^i_{RR}C^{i*}_{LR}
                  -C^i_{LL}C^{i*}_{RL}\right]\,,      \nonumber\\
&& C^{i\prime}_{3}=\frac{1}{4}
             \left[|C^i_{RR}|^2+|C^i_{LR}|^2
                  -|C^i_{RL}|^2-|C^i_{LL}|^2\right]\,.
\label{eq:decay quartic charge}
\end{eqnarray}
The polarized distribution with a polarization vector $\bar{n}_j^{\mu}$ 
of the neutralino decay
$\tilde{\chi}^0_i\rightarrow\tilde{\chi}^0_1\,\ell^{\prime -}\ell^{\prime +}$
can be derived in a straightforward way
\begin{eqnarray}
|\bar{\cal D}|^2(\bar{n}_j) &=& 
 -4(t''- m_{\tilde{\chi}^0_j}^2)
   (t''- m_{\tilde{\chi}^0_1}^2)(N^j_1+N^j_3)
 -4(u''- m_{\tilde{\chi}^0_j}^2)
   (u''- m_{\tilde{\chi}^0_1}^2)(N^j_1-N^j_3)   \nonumber\\
&& -8 m_{\tilde{\chi}^0_j} m_{\tilde{\chi}^0_1} s''N^j_2 \nonumber \\
&& -8 (\bar{n}_j\cdot\bar{q}')\left[
-m_{\tilde{\chi}^0_j}(m_{\tilde{\chi}^0_1}^2-u'')(N^{j\prime}_1-N^{j\prime}_3)
+m_{\tilde{\chi}^0_1}(m_{\tilde{\chi}^0_j}^2-t'')N^{j\prime}_2\right]
    \nonumber\\
&& +8 (\bar{n}_j\cdot q')\left[
 m_{\tilde{\chi}^0_j}(m_{\tilde{\chi}^0_1}^2-t'')(N^{j\prime}_1+N^{j\prime}_3)
+m_{\tilde{\chi}^0_1}(m_{\tilde{\chi}^0_j}^2-u'')N^{j\prime}_2\right]
    \nonumber\\
&& +16\, m_{\tilde{\chi}^0_1}\langle q_j \bar{n}_j q' \bar{q}'\rangle N^j_4\,,
\label{eq:decay dist-antiptl}
\end{eqnarray}
The quartic charges for the neutralino decay case can be obtained from
the bilinear charges $N^j_{\alpha\beta}$ in the same way as
the chargino quartic charges
are defined in terms of the bilinear charges $C^i_{\alpha\beta}$.
Related to CP violation it is worthwhile to note that
the quartic charges $C^i_4$ and $N^j_4$ manifest
CP violation in the theory.\\

For the sake of subsequent discussion of the spin/angular correlations between 
the production and decay processes, we construct the decay 
density matrix $\rho_{\lambda\lambda'}\sim {\cal D}_\lambda {\cal 
D}^*_{\lambda'}$. In general, the decay amplitude for a spin--1/2 particle 
and its complex conjugate can be expressed as 
\begin{eqnarray}
{\cal D}(\lambda) = \Gamma\, u(q,\lambda),\qquad 
{\cal D}^*(\lambda') = \bar{u}(q,\lambda')\,\bar{\Gamma},
\end{eqnarray}
with the general spinor structure $\Gamma$ and $\bar{\Gamma}=\gamma^0
\Gamma^\dagger$. Then we use the general formalism to calculate
the decay density matrix involving a particle with 4--momentum $q$ and 
mass $m$ by introducing three spacelike 4--vectors 
$n^a_\mu$ ($a=1,2,3$) which together with $q/m\equiv n^0$ form an orthonormal 
set:
\begin{eqnarray}
g^{\mu\nu}\, n^a_\mu n^b_\nu =g^{ab},\qquad 
g_{ab}\, n^a_\mu n^b_\nu = g_{\mu\nu}\,,
\end{eqnarray}
with $g^{\mu\nu},\,g^{ab}={\rm diag}\,(1,-1,-1,-1)$ $(a,b=0,1,2,3)$. 
A convenient choice for the explicit 
form of ${n^a}^\mu$ is in a coordinate system where the direction of the 
three--momentum of the particle is $\hat{q}=(\sin\theta,0,\cos\theta)$ 
lying on the $x$-$z$ plane:
\begin{eqnarray}
{n^1}^\mu=(0,\cos\theta,0,-\sin\theta)\,,\qquad
{n^2}^\mu=(0,0,1,0)\,,\qquad 
{n^3}^\mu=\frac{1}{m}\left(|\vec{q}|,E\hat{q}\right)\,.
\end{eqnarray}
Then in the given reference frame, the three 4--vectors as defined in the
above equation describe the transverse, normal
and longitudinal polarization of the decaying particle, respectively.\\

With the four--dimensional orthonormal basis of the 4--vectors 
$\{n^0,n^1,n^2,n^3\}$,
we can derive the so--called Bouchiat--Michel formulas \cite{BM}
for $u$ and $v$ spinors
\begin{eqnarray}
u(q,\lambda)\bar{u}(q,\lambda')=\frac{1}{2}
   \left[\delta_{\lambda\lambda'}+\gamma_5\not\!{n}^a\tau^a_{\lambda'\lambda}
   \right](\not\!{q}+m)\,, \nonumber \\
v(q,\lambda)\bar{v}(q,\lambda')=\frac{1}{2}
   \left[\delta_{\lambda\lambda'}+\gamma_5\not\!{n}^a\tau^a_{\lambda\lambda'}
   \right](\not\!{q}-m)\,,
\label{eq:B-M}
\end{eqnarray}
with $\lambda,\lambda'=\pm$.
These formulas enable us to compute the squared, normalized 
decay density matrix $\rho_{\lambda\lambda'}$ as follows:
\begin{eqnarray}
\rho_{\lambda\lambda'}\equiv
      \frac{{\cal D}(\lambda){\cal D}^*(\lambda')}{\sum_{\lambda}
            |{\cal D}(\lambda)|^2} = \frac{1}{2}\left[\delta_{\lambda\lambda'}
           +\frac{Y^{a}}{X}\tau^{a}_{\lambda'\lambda}\right]
\end{eqnarray}
where $\tau^a$ ($a=1,2,3$) are the Pauli matrices.
The three functions $X$ and $Y^a$ ($a=1,2,3$)  
for the chargino leptonic decay $\tilde{\chi}^-_i\rightarrow
\tilde{\chi}^0_1\, \ell^-\bar{\nu}_\ell$ and the three functions $\bar{X}$ and 
$\bar{Y}^a$ ($a=1,2,3$) for the neutralino leptonic decay 
$\tilde{\chi}^0_j\rightarrow\tilde{\chi}^0_1\,\ell^{\prime +}
\ell^{\prime -}$ 
can be obtained easily as 
\begin{eqnarray}
X &=& -8(t'- m_{\chi_i^-}^2)(t'- m_{\chi_1^0}^2)(C^i_1-C^i_3) 
      -8(u'- m_{\chi_i^-}^2)(u'- m_{\chi_1^0}^2)(C^i_1+C^i_3)   \nonumber\\
 && -16\, m_{\tilde{\chi}^-_i} m_{\tilde{\chi}^0_1} s'C^i_2, \nonumber \\
Y^a &=&-16 (n_i^a\cdot\bar{q})\left[
  m_{\tilde{\chi}^-_i}(m_{\tilde{\chi}^0_1}^2-u')(C^{i\prime}_1+C^{i\prime}_3)
 +m_{\tilde{\chi}^0_1}(m_{\tilde{\chi}^-_i}^2-t')C^{i\prime}_2\right]
      \nonumber\\
 && +16(n_i^a\cdot q)\left[
 -m_{\tilde{\chi}^-_i}(m_{\tilde{\chi}^0_1}^2-t')(C^{i\prime}_1-C^{i\prime}_3)
 +m_{\tilde{\chi}^0_1}(m_{\tilde{\chi}^-_i}^2-u')C^{i\prime}_2\right]
      \nonumber\\ 
 && -32\, m_{\tilde{\chi}^0_1}\langle q_i n_i^a q \bar{q}\rangle C^i_4\,,
\end{eqnarray}
and 
\begin{eqnarray}
\bar{X} &=& -8(t''- m_{\tilde{\chi}^0_j}^2)
              (t''- m_{\tilde{\chi}^0_1}^2)(N^j_1+N^j_3)
            -8(u''- m_{\tilde{\chi}^0_j}^2)
              (u''- m_{\tilde{\chi}^0_1}^2)(N^j_1-N^j_3)   \nonumber\\
 && -16\, m_{\tilde{\chi}^0_j} m_{\tilde{\chi}^0_1} s''N^j_2, \nonumber \\
\bar{Y}^a &=&-16 (\bar{n}_j^a\cdot\bar{q}')\left[
 -m_{\tilde{\chi}^0_j}(m_{\tilde{\chi}^0_1}^2-u'')(N^{j\prime}_1-N^{j\prime}_3)
 +m_{\tilde{\chi}^0_1}(m_{\tilde{\chi}^0_j}^2-t'')N^{j\prime}_2\right]
     \nonumber\\
&& +16(\bar{n}_j^a\cdot q')\left[
 +m_{\tilde{\chi}^0_j}(m_{\tilde{\chi}^0_1}^2-t'')(N^{j\prime}_1+N^{j\prime}_3)
 +m_{\tilde{\chi}^0_1}(m_{\tilde{\chi}^0_j}^2-u'')N^{j\prime}_2\right]
     \nonumber\\
 && +32\, m_{\tilde{\chi}^0_1}\langle q_j \bar{n}_j^a q' \bar{q}'\rangle N^j_4\,,
\end{eqnarray}
where $n_i^a$ and $\bar{n}_j^a$ is the polarization vector of the decaying 
chargino and neutralino, respectively.\\

\subsection{Branching fractions}
\label{subsec:5.2}

The main decay modes of the lightest chargino $\tilde{\chi}^-_1$ and 
the next--lightest neutralino $\tilde{\chi}^0_2$ can be classified as follows:
\begin{eqnarray}
 \tilde{\chi}^-_1&\rightarrow& W^*\tilde{\chi}^0_1,\, H^{*}\tilde{\chi}^0_1
                    \rightarrow \tilde{\chi}^0_1\,\ell^-\bar{\nu}_\ell,\,
                                \tilde{\chi}^0_1 q\bar{q'}\,,\nonumber\\
 \tilde{\chi}^-_1&\rightarrow& \ell\tilde{\nu}^*,\,
                                 \nu\tilde{\ell}^*,\, q\tilde{q'}^*
                    \rightarrow \tilde{\chi}^0_1\,\ell^-\bar{\nu}_\ell,\,
                                \tilde{\chi}^0_1 q\bar{q'}\,,\nonumber\\
 \tilde{\chi}^0_2&\rightarrow& Z^*\tilde{\chi}^0_1,\, H^*\tilde{\chi}^0_1
                    \rightarrow \tilde{\chi}^0_1\,\ell^+\ell^-,\,
                                \tilde{\chi}^0_1 q\bar{q}\,,\nonumber\\
 \tilde{\chi}^0_2&\rightarrow& \ell\tilde{\ell}^*,\nu\tilde{\nu}^*,\,
                                 q\tilde{q}^*
                    \rightarrow  \tilde{\chi}^0_1\,\ell^+\ell^-,\,
                                 \tilde{\chi}^0_1 q\bar{q}\,,
\end{eqnarray}
with $q$ and $q'$ belonging to the same SU(2)$_L$ multiplet.
Besides, if the mass $m_{\tilde{\chi}^\pm_1}$ is smaller than the
neutralino mass $m_{\tilde{\chi}^0_2}$, the lightest chargino 
$\tilde{\chi}^\pm_1$ can take part in the neutralino decay via
the processes $\tilde{\chi}^0_2\rightarrow\tilde{\chi}^\pm_1 W^{\mp *},\,
\tilde{\chi}^\pm_1 H^{\mp *}$ and vice versa. 
Concerning the main decay modes, there are several aspects 
to be noted:
\begin{itemize}
\item For the first and second generation leptons, the Higgs--exchange
      diagrams are suppressed if $\tan\beta$ is not very large and
      there is no generational mixing in the slepton sector.
\item The experimental bounds on the Higgs particles are very stringent
      so that the two--body decays $\tilde{\chi}^0_2\rightarrow 
      H\tilde{\chi}^0_1$, $\tilde{\chi}^0_2\rightarrow H^\pm
      \tilde{\chi}^\mp_1$, and $\tilde{\chi}^-_1\rightarrow
      H^-\tilde{\chi}^0_1$ are expected to be not available or at least
      strongly suppressed. If these decay modes open up, it could then spoil
      the tri--lepton signal. 
\item The lightest chargino and the second--lightest neutralino are
      almost degenerate in the gaugino--dominated parameter space  
      so that the charged decays such as $\tilde{\chi}^0_2
      \rightarrow\tilde{\chi}^\pm_1\,\ell^\mp\nu_\ell$ and $\tilde{\chi}^-_1
      \rightarrow\tilde{\chi}^0_2\,\ell^-\bar{\nu}_\ell$ will be highly 
      suppressed.
\end{itemize}
Keeping in mind the above subtle aspects, we calculate the branching fractions 
${\cal B}(\tilde{\chi}^-_1\rightarrow\tilde{\chi}^0_1\,\ell^-\bar{\nu}_\ell)$ and 
${\cal B}(\tilde{\chi}^0_2\rightarrow\tilde{\chi}^0_1\,\ell^{\prime +}
\ell^{\prime -})$ fully incorporating all the possible decay modes of the 
neutralino $\tilde{\chi}^0_2$ but neglecting the Higgs-exchange contributions
in both scenarios.\\

We present in Fig.~\ref{fig:fig10} the branching fractions 
${\cal B}(\tilde{\chi}^-_1\rightarrow\tilde{\chi}^0_1\,\ell^-\bar{\nu}_\ell)$ 
and ${\cal B}(\tilde{\chi}^0_2\rightarrow\tilde{\chi}^0_1 
\,\ell^{\prime +}\ell^{\prime -})$ for $\ell\,\ell^\prime =e$ or $\mu$ in 
${\cal S}1$ (two upper figures) and in ${\cal S}2$ (two lower figures). 
In the scenario ${\cal S}1$ the branching fraction
${\cal B}(\tilde{\chi}^-_1\rightarrow\tilde{\chi}^0_1\,\ell^-\bar{\nu}_\ell)$ is 
almost constant over the whole space of the phases and
the branching fraction ${\cal B}(\tilde{\chi}^0_2\rightarrow\tilde{\chi}^0_1
\,\ell^{\prime +}\ell^{\prime -})$ is very small and sensitive to the phases
only around 
$\Phi_1$ around 
$\Phi_\mu=0,2\pi$. 
The insensitivity of both branching fractions to the phases
in ${\cal S}1$ is due to the fact that
the $t$-- and $u$--channel contributions are suppressed due to large 
slepton masses and the couplings ${\cal W}_{L11}, {\cal W}_{R11}$, and 
${\cal Z}_{21}$ for the $s$--channel contributions are not so much 
sensitive to $\Phi_{\mu}$ and $\Phi_1$. On the contrary, the branching 
fractions are rather sensitive to $\Phi_\mu$ and $\Phi_1$ in the 
scenario ${\cal S}2$. 
It is interesting that the branching fraction ${\cal B}(\tilde{\chi}^-_1
\rightarrow\tilde{\chi}^0_1\,\ell^-\bar{\nu}_\ell)$ can be minimal for certain
non--trivial values of the CP--violating phases. 
We note that the branching fraction ${\cal B}(\tilde{\chi}^0_2\rightarrow
\tilde{\chi}^0_1\,\ell^{\prime +}\ell^{\prime -})$ is greatly enhanced in 
the scenario ${\cal S}2$; this stems from the fact that the slepton--exchange 
contributions 
due to mainly the gaugino components of the neutralinos become dominant 
for the small slepton masses and the large value of $|\mu|$. 
On the contrary, the branching fraction ${\cal B}(\tilde{\chi}^-_1\rightarrow
\tilde{\chi}^0_1\,\ell^-\bar{\nu}_\ell)$ is not so different in size between the
two scenarios, but the branching fraction becomes more sensitive to the
CP--violating phases in the scenario ${\cal S}2$.\\

In summary, the branching fractions ${\cal B}(\tilde{\chi}^-_1
\rightarrow \tilde{\chi}^0_1\,\ell^-\bar{\nu}_\ell)$ and 
${\cal B}(\tilde{\chi}^0_2\rightarrow \tilde{\chi}^0_1\,\ell^{\prime +}
\ell^{\prime -})$ are not so strongly dependent
on the CP--violating phases $\{\Phi_\mu,\Phi_1\}$, but 
the branching fraction ${\cal B}(\tilde{\chi}^0_2\rightarrow\tilde{\chi}^0_1
\,\ell^{\prime +}\ell^{\prime -})$ is very sensitive to the slepton masses.

\section{Spin/Angular Correlated Observables}
\label{sec:correlation}

\subsection{Correlations between production and decay}
\label{subsec:6.1}

In this section we provide a general formalism to describe the
spin/angular correlations between the production 
process $d\bar{u}\rightarrow\tilde{\chi}^-_i\tilde{\chi}^0_j$
and the sequential leptonic decays of $\tilde{\chi}^-_i$
and $\tilde{\chi}^0_j$.
Formally  we can have the spin/angular correlated distribution by 
taking the sum over the helicity indices of the intermediate 
chargino and neutralino states and folding with the chargino--neutralino
leptonic
decay density matrix $\rho_{\lambda\lambda'}$ and  
$\bar{\rho}_{\bar{\lambda}\bar{\lambda'}}$
with the matrix squared for the production helicity amplitudes:  
\begin{eqnarray}
\sum_{\rm corr} &\equiv& \pi^2\alpha^2
     \sum_{\lambda\lambda'}\sum_{\bar{\lambda}\bar{\lambda'}}\sum_\sigma
     \langle\sigma;\lambda\bar{\lambda}\rangle
     \langle\sigma;\lambda'\bar{\lambda'}\rangle^*
     \rho_{\lambda\lambda'}
     \bar{\rho}_{\bar{\lambda}\bar{\lambda'}}          \nonumber \\
  &=& \pi^2 \alpha^2 \bigg[\Sigma_{\rm unp}
    +P_z{\cal P}+\bar{P}_z\bar{{\cal P}}
    +P_x{\cal U}+P_y\bar{{\cal U}}
    +\bar{P}_x{\cal V}+\bar{P}_y\bar{{\cal V}}+P_z\bar{P}_z{\cal Q}\nonumber\\
  &&\hskip 0.4cm +P_x\bar{P}_z{\cal W}
    +P_y\bar{P}_z\bar{{\cal W}}
    +\bar{P}_x P_z{\cal X}+\bar{P}_y P_z\bar{{\cal X}} 
    +(P_x\bar{P}_x-P_y\bar{P}_y){\cal Y} \nonumber\\
  &&\hskip 0.4cm +(P_x\bar{P}_y+P_y\bar{P}_x)\,\bar{{\cal Y}}       
    +(P_x\bar{P}_x+P_y\bar{P}_y)\,{\cal Z}
    +(P_y\bar{P}_x-P_x\bar{P}_y)\,\bar{{\cal Z}}\bigg],
\label{eq:spin correlation}
\end{eqnarray}
where the functions $P_{x,y,z}$ and $\bar{P}_{x,y,z}$ depending on the 
chargino and neutralino decay distributions, respectively,  
\begin{eqnarray}
&& P_x = \frac{Y^1}{X}\,,\qquad
   P_y = \frac{Y^2}{X}\,,\qquad
   P_z = \frac{Y^3}{X}\,,\nonumber\\[1mm] 
&& \bar{P}_x = \frac{\bar{Y}^1}{\bar{X}}\,,\qquad
   \bar{P}_y = \frac{\bar{Y}^2}{\bar{X}}\,,\qquad
   \bar{P}_z = \frac{\bar{Y}^3}{\bar{X}}\,, 
\end{eqnarray}
serve as the polarimeters to extract the spin--spin correlations of the
chargino $\tilde{\chi}^-_i$ and the neutralino $\tilde{\chi}^0_j$
in the production process.
The sixteen coefficients are combinations of the
production helicity amplitudes,
corresponding
to the unpolarized cross section, $2 \times 3$ polarization components and
$3 \times 3$ spin--spin correlations.

\bigskip

(i) \underline{\bf Unpolarized part}:
\begin{eqnarray}
\Sigma_{\rm unpol}&=&\frac{1}{4}\sum_{\sigma=\pm}
      \bigg[|\langle\sigma;++\rangle|^2+|\langle\sigma;+-\rangle|^2
           +|\langle\sigma;-+\rangle|^2+|\langle\sigma;--\rangle|^2
      \bigg]\,.
\end{eqnarray}

\smallskip

(ii) \underline{\bf Polarization components}:

\begin{eqnarray}
{\cal P}&=&\frac{1}{4}\sum_{\sigma=\pm}
      \bigg[|\langle\sigma;++\rangle|^2+|\langle\sigma;+-\rangle|^2
           -|\langle\sigma;-+\rangle|^2-|\langle\sigma;--\rangle|^2
      \bigg]\,, \nonumber\\
\bar{\cal P}&=&\frac{1}{4}\sum_{\sigma=\pm}
      \bigg[|\langle\sigma;++\rangle|^2+|\langle\sigma;-+\rangle|^2
           -|\langle\sigma;+-\rangle|^2-|\langle\sigma;--\rangle|^2
      \bigg]\,, \nonumber\\
{\cal U}&=&\frac{1}{2}\sum_{\sigma=\pm}
      \real\bigg\{\langle\sigma;-+\rangle\langle\sigma;++\rangle^*
                    +\langle\sigma;--\rangle\langle\sigma;+-\rangle^*\bigg\}
       \,,\nonumber\\
{\cal V}&=&\frac{1}{2}\sum_{\sigma=\pm}
      \real\bigg\{\langle\sigma;+-\rangle\langle\sigma;++\rangle^*
                    +\langle\sigma;--\rangle\langle\sigma;-+\rangle^*\bigg\}
      \,,
\end{eqnarray}
and $\bar{\cal U}, \bar{\cal V}$ defined as ${\cal U}, {\cal V}$
after replacing $\real$ by $\imag$. 
We note in passing that the above combinations are directly related with the 
polarization vector of the chargino $\tilde{\chi}^-_1$ and 
neutralino $\tilde{\chi}^0_2$ defined in Sect.~\ref{subsec:4.4}
\bigskip

(iii) \underline{\bf Spin--spin correlations}:
\begin{eqnarray}
{\cal Q}&=&\frac{1}{4}\sum_{\sigma=\pm}
      \bigg[|\langle\sigma;++\rangle|^2-|\langle\sigma;+-\rangle|^2
           -|\langle\sigma;-+\rangle|^2+|\langle\sigma;--\rangle|^2
      \bigg]\,,\nonumber\\
{\cal W}&=&\frac{1}{2}\sum_{\sigma=\pm}
      \real\bigg\{\langle\sigma;-+\rangle\langle\sigma;++\rangle^*
                    -\langle\sigma;--\rangle\langle\sigma;+-\rangle^*\bigg\}
       \,,\nonumber\\
{\cal X}&=&\frac{1}{2}\sum_{\sigma=\pm}
      \real\bigg\{\langle\sigma;+-\rangle\langle\sigma;++\rangle^*
                    -\langle\sigma;--\rangle\langle\sigma;-+\rangle^*\bigg\}
       \,,\nonumber\\
{\cal Y}&=&\frac{1}{2}\sum_{\sigma=\pm}
      \real\bigg\{\langle\sigma;--\rangle\langle\sigma;++\rangle^*\bigg\}
       \,,\nonumber\\
{\cal Z}&=&\frac{1}{2}\sum_{\sigma=\pm}
      \real\bigg\{\langle\sigma;-+\rangle\langle\sigma;+-\rangle^*\bigg\}
      \,,
\end{eqnarray}
and $\bar{\cal W}, \bar{\cal X}, \bar{\cal Y}, \bar{\cal Z}$ defined
as ${\cal W}, {\cal X}, {\cal Y}, {\cal Z}$ after replacing $\real$
by $\imag$, and  these components along with $\bar{\cal U}$ and $\bar{\cal V}$ 
will contribute to CP--odd observables. \\

Combining the production and decay distributions, we obtain the 
fully spin/angular correlated 11--fold differential cross section 
for the parton--level process $d\bar{u}\rightarrow
\tilde{\chi}^-_i\tilde{\chi}^0_j\rightarrow (\tilde{\chi}^0_1\,\ell^-
\bar{\nu}_\ell) (\tilde{\chi}^0_1\,\ell^{\prime -}\ell^{\prime +})$:
\begin{eqnarray}
{\rm d}\sigma=\frac{\pi\alpha^2\beta}{8s}\,
         {\cal B}(\tilde{\chi}^-_i\rightarrow\tilde{\chi}^0_1
                  \,\ell^-\bar{\nu}_\ell) \,
         {\cal B}(\tilde{\chi}^0_j\rightarrow\tilde{\chi}^0_1
         \,\ell^{\prime -}\ell^{\prime +}) \,\,
          \sum_{\rm corr}\,\, {\rm d}\Phi_{3\ell}\,,
\label{eq:correlated}
\end{eqnarray}
where ${\rm d}\Phi_{3\ell}$ denotes the final--state phase space volume
element and it can be parameterized in terms of 11 independent kinematical
variables as follows:
\begin{eqnarray*}
{\rm d}\Phi_{3\ell}= {\rm d}\cos\Theta\, 
      {\rm d}x_1\,{\rm d}x_2\,{\rm d}\cos\theta_1\,{\rm d}\phi_1\,
      {\rm d}\phi_{12}\, {\rm d}x_3\,{\rm d}x_4\,{\rm d}\cos\theta_3\,
      {\rm d}\phi_3\,{\rm d}\phi_{34}\,,
\end{eqnarray*}
where the angular variable $\theta_1$ is the polar 
angle of the 
$\ell^-$ in the $\tilde{\chi}^-_i$ rest frame with respect to the original 
flight direction in the parton--level $d\bar{u}$ center of mass  frame, 
and $\phi_1$ the 
corresponding azimuthal angle with respect to the production plane, and 
$\phi_{12}$ is the relative azimuthal angle of $\bar{\nu}_\ell$ 
along the $\ell^-$ direction with respect to the production plane.
A similar configuration can be specified for the neutralino decay distribution
by $\theta_3$, $\phi_3$ and $\phi_{34}$. 
The dimensionless parameters $x_1$, $x_2$, $x_3$, and $x_4$ denote 
the lepton energy fractions
\begin{eqnarray}
x_1 = \frac{2E_{\ell^-}}{m_{\tilde{\chi}^-_i}}\,,\qquad
x_2 = \frac{2E_{\nu}}{m_{\tilde{\chi}^-_i}}\,,\qquad
x_3 = \frac{2E_{\ell^{\prime -}}}{m_{\tilde{\chi}^0_j}}\,,\qquad
x_4 = \frac{2E_{\ell^{\prime +}}}{m_{\tilde{\chi}^0_j}}\,.
\end{eqnarray}
The allowed space of the kinematical variables for the chargino leptonic
decay is determined by the kinematic conditions obtained with  
the masses of the final--state leptons neglected:
\begin{eqnarray}
&& 0\leq \Theta    \leq \pi;\qquad  0\leq \theta_1  \leq \pi\,, \ \
   0\leq \phi_1    \leq 2\pi\,,\ \
   0\leq \phi_{12} \leq 2\pi; \nonumber\\
&& 0\leq x_{1,2}   \leq 1-r_{i1}\,,   \ \
   (1-x_1)(1-x_2)\geq r_{i1} \,, \ \
   x_1+x_2\geq 1-r_{i1}\,,
\end{eqnarray}
where $r_{i1}=m^2_{\tilde{\chi}^0_1}/m^2_{\tilde{\chi}^-_i}$,
and similarly the allowed range of the kinematical observables for
the neutralino leptonic decay can be obtained simply by replacing
the labels; $(1,2)$ to $(3,4)$ and $r_{i1} \to r_{j1}=
m^2_{\tilde{\chi}^0_1}/m^2_{\tilde{\chi}^0_j}$.\\

Finally,\, the tri--lepton rates in the $p\bar{p}$ 
laboratory frame is obtained by folding the parton--level differential
cross section (\ref{eq:correlated})
with the $d$-quark and $\bar{u}$-quark parton distributions.
This folding involves 2 additional kinematical variables $\tau$ and $x$. 
Since the parton--level c.m. frame is not fixed with respect to the
$p\bar{p}$ c.m. frame the Lorentz boost of the partonic system along
the initial beam direction has to be properly taken into account.

\subsection{Total cross section of the correlated process}
\label{subsec:6.2}

The total cross section for the correlated process $p\bar{p}\rightarrow 3l+ X$ 
can be obtained simply by computing 
\begin{eqnarray}
\sigma(p\bar{p}\rightarrow 3\ell+X)
 =\sigma(p\bar{p}\rightarrow\tilde{\chi}^-_1\tilde{\chi}^0_2)\, 
  {\cal B}(\tilde{\chi}^-_i\rightarrow\tilde{\chi}^0_1\,\ell^-\bar{\nu}_\ell)\,
  {\cal B}(\tilde{\chi}^0_2\rightarrow\tilde{\chi}^0_1
           \,\ell^{\prime +}\ell^{\prime -})\,.
\end{eqnarray}

We present in Fig.~\ref{fig:fig11} the contour plots for the total cross 
section on the plane of two CP--violating phases $\{\Phi_\mu,\Phi_1\}$ in 
the two scenarios (a) ${\cal S}1$ and (b) ${\cal S}2$.
In the scenario ${\cal S}1$ the branching fraction ${\cal B}(\tilde{\chi}^-_1
\rightarrow\tilde{\chi}^0_1\,\ell^-\bar{\nu}_\ell)$ is almost constant as 
shown in Fig.~\ref{fig:fig10} so that the total tri--lepton cross section 
is mainly determined by 
${\cal B}(\tilde{\chi}^0_2\rightarrow\tilde{\chi}^0_1\,\ell^{\prime +}
\ell^{\prime -})$ 
and the production cross section $\sigma(p\bar{p}\rightarrow\tilde{\chi}^-_1
\tilde{\chi}^0_2)$.  
The total tri--lepton cross section is of the order of $1$~fb (10 fb) for the
parameter sets ${\cal S}1$ (${\cal S}2$), respectively. However in the 
future Tevatron RUN-II
experiments with the upgraded luminosity of the order of 2 fb$^{-1}$  
a handful of tri--lepton events can be produced. 
Note that the total cross section is sensitive to the 
CP-violating phases in the scenario ${\cal S}2$ 
and the cross section can be minimal for certain non--trivial values of
the CP--violating phases. This property is mainly due to the fact that
the branching fraction ${\cal B}(\tilde{\chi}^-_1\rightarrow\tilde{\chi}^0_1
\,\ell^-\bar{\nu}_\ell)$ exhibits a similar pattern as shown in 
Fig.~\ref{fig:fig10}.
Therefore, depending on the size of the integrated luminosity
the very existence of the minimum event rate and the simultaneous small 
mass splitting (See Fig.~\ref{fig:fig4}) for the non--trivial CP--violating 
phases reflect that the Tevatron bounds on $m_{\tilde{\chi}^\pm_1}$ and
$m_{\tilde{\chi}^0_2}$ might be much smaller than those \cite{CCEFM} 
ruled out in the context of SUGRA and GUT inspired SUSY models.

\subsection{Dilepton invariant mass distributions}
\label{subsec:6.3}

The final--state leptons of the neutralino decay 
$\tilde{\chi}^0_2\rightarrow\tilde{\chi}^0_1\, \ell^{\prime +}
\ell^{\prime -}$ provides us with a very easily measurable kinematical 
observable; the dilepton invariant mass, $m_{\ell\ell}$. 
This Lorentz--invariant quantity can be precisely 
reconstructed by measuring the two lepton momenta, and it is 
nothing but the square root of the Mandelstam variable, $\sqrt{s''}$,
\begin{eqnarray}
m_{\ell\ell}=\sqrt{s''}=m_{\tilde{\chi}^0_j}\sqrt{x_3+x_4-1+r_{j1}}\,.
\end{eqnarray}
Furthermore, the invariant mass distribution is independent of the specific 
production process for the parent neutralino $\tilde{\chi}^0_2$, because 
the invariant mass does not involve any angular variables describing 
the decays so that the polarization of the decaying neutralino
does not affect the distribution.\\

Figure~\ref{fig:fig12} shows the dilepton invariant mass distribution
in the scenarios (a) ${\cal S}1$ and (b) ${\cal S}2$. 
In principle, three dilepton combinations  
can be constructed out of the three charged leptons.
But, the correct dilepton combination 
will have a sharp end point of its invariant mass distribution 
with a distinguishable peak in the dilepton invariant mass distribution. 
This allows one to experimentally determine
the mass difference $\Delta_m=m_{\tilde{\chi}^0_2}-m_{\tilde{\chi}^0_1}$
with a good precision. Note that the position of the end points is strongly 
dependent on the CP--violating phases.  
It can provide us with the opportunity to probe the CP violating 
phases if all the other real parameters are known.
As the gaugino masses vary 
significantly with the relevant CP violating phases, the 
distributions with respect to the kinematic variables like the energies and 
transverse momenta of the final--state leptons are strongly
influenced by those CP--violating phases. 

\subsection{Lepton angular distribution in the laboratory frame}
\label{subsec:6.4}

In a realistic experimental situation, 
it is necessary to isolate tri--lepton signal events by applying 
various selection cuts effectively to reduce the contaminations 
from all the  background processes as much as possible. 
For that purpose, it is very important to fully understand the event 
topology of tri--lepton signal which depends on the polarizations of the  
chargino and neutralino at the intermediate stage.
The full incorporation of the spin--correlations (\ref{eq:spin correlation}) 
involves a lot of correlated terms, which many Monte Carlo 
simulations \cite{MOST} have simply neglected by including only the first 
un--correlated term in eq.~(\ref{eq:spin correlation}). In this
section, we take an 
easily--measurable kinematical variable, the scattering angle $\theta_{\ell}$
of the lepton from the chargino decay with respect to the proton beam
direction and estimate the variation of the lepton 
angular distribution of the tri--lepton signal due to the spin--correlation 
effects.\\

Figure~\ref{fig:fig13} exhibits the lepton angular distribution of the 
tri--lepton signature for both the un--correlated case and the 
fully--correlated case for different phase combinations
in both scenarios.
Note that the correlated distribution can be different from the
the un--correlated distribution around the forward and backward regions
depending on the values of the CP--violating phases. In particular,
the forward--backward asymmetry can be different in two cases.
Therefore, even for this simple distribution it is important to 
take into account the spin correlations between production and decay 
fully to obtain the magnitude properly. Without any experimental cuts, 
such observables as the total tri--lepton event rate can be independent 
of whether the spin/angular correlations are taken into account or not. 
However, it might be inevitable to apply some efficient experimental cuts 
to suppress serious backgrounds. In that case the spin--correlations should 
be included.

\subsection{CP--odd triple momentum products}
\label{subsec:6.5}

So far we have concentrated mainly on the CP--even production--decay
correlated observables which depend on the CP phases only indirectly. 
For direct measurements of the CP--violating phases one has to 
use CP--odd or T-odd observables.  Some of such CP--odd 
(or T--odd) observables can be constructed by taking a triple product of 
any combination of the initial proton (or anti--proton) momentum and the three
final lepton momenta. One typical example is the following triple momentum 
product (TMP):
\begin{eqnarray}
{\cal O}_T=\vec{p}_{_{\ell_1}}\cdot (\vec{p}_{_{\ell_3}}
           \times\vec{p}_{_{\ell_4}}),
\label{eq:T-odd observable1}
\end{eqnarray}
where $\ell_1=\ell^-$ of the chargino decay $\tilde{\chi}^-_1\rightarrow
\tilde{\chi}^0_1\,\ell^-\bar{\nu}_\ell$, and $\ell_3=\ell^{\prime -}$,
$\ell_4=\ell^{\prime +}$ of the neutralino decay $\tilde{\chi}^0_2
\rightarrow \tilde{\chi}^0_1\,\ell^{\prime -}\ell^{\prime +}$.
The observable (\ref{eq:T-odd observable1}) enables us to probe 
the CP--violating phases directly when neglecting the tiny 
particle decay widths. Similarly, the initial proton momentum and two 
final--state leptons allows us to construct additional T--odd observables:
\begin{eqnarray}
{\cal O}^{\ell\ell'}_T=\vec{p}_p\cdot\left(\vec{p}_\ell\times
          \vec{p}_{\ell'}\right)
\label{eq:T-odd observable2}
\end{eqnarray}
where $\{\ell,\ell'\}$ is any combination of two momenta among the three final 
lepton momenta. In total, we have four independent TMPs;
${\cal O}_T$, ${\cal O}^{\ell_1\ell_3}_T$, ${\cal O}^{\ell_1\ell_4}_T$ and
${\cal O}^{\ell_3\ell_4}_T$.\\

In general, any of T--odd  
TMP can be given by a linear combination of the 
quartic charges $\{Q^{ij}_4, C^i_4,N^j_4\}$. We note that the same topological
pattern of the contributing diagrams between the associated production and 
the chargino decay make $Q^{11}_4$ and $C^1_4$ correlated in size.  
Even before making any numerical
estimate of the T--odd triple products, we can argue that their size is
very small in the scenario ${\cal S}1$ with heavy sfermion masses. 
Firstly, the chargino and neutralino normal polarizations
are very small in the scenario ${\cal S}1$ as shown in Sect.~\ref{subsec:4.4},
implying small $Q^{12}_4$ and $C^1_4$. Secondly, with the
negligible $t$-- and $u$--channel slepton contributions, 
the remaining quartic charge $N_4$ of the neutralino $\tilde\chi_2^0$ 
simplifies to the expression
\begin{eqnarray}
N_4 \approx \frac{|D_Z^{''}|^2}{c^4_W s^4_W}\,\left(s^2_W-\frac{1}{4}\right)\,
           \imag\left({\cal Z}^2_{21}\right)\,,
\end{eqnarray} 
which contains a very small numerical factor $(s^2_W-1/4)\sim -0.02$ 
\cite{PDG98}. Therefore, the quartic charge $N_4$ is also extremely 
suppressed in the scenario ${\cal S}1$. This simultaneous suppression of
three quartic charges render all the TMPs strongly suppressed 
in the scenario ${\cal S}1$. \\

The three TMPS $\{{\cal O}_T,\, {\cal O}^{\ell_1\ell_3}_T,\, 
{\cal O}^{\ell_1\ell_4}_T\}$ 
involve both the chargino and neutralino leptonic decays, but the TMP 
${\cal O}^{\ell_3\ell_4}$ involves only the neutralino leptonic decay. 
Therefore, one can have a large statistical gain by exploiting the observable
${\cal O}^{\ell_3\ell_4}_T$ in measuring the CP--violating phases.
In this light, we consider only the observable  ${\cal O}^{\ell_3\ell_4}_T$ 
to probe the CP--odd phases which are not excluded by
the EDM constraints. 
Certainly, one needs to estimate all the possible systematic uncertainties 
in determining the reconstruction efficiency of the 
$\tilde{\chi}^\pm_1\tilde{\chi}^0_2$ mode. Nevertheless, 
let us take into account only the statistical errors including the
hadronic decay modes of the chargino $\tilde{\chi}^-_i$ in the present work
in which case the excluded region of $\{\Phi_\mu,\Phi_1\}$ at the 
$N$--$\sigma$ level for a given 
integrated luminosity $\int {\cal L}\,{\rm d}t$ satisfies the inequality:
\begin{eqnarray}
\int {\cal L}\, {\rm d}t\geq\frac{N}{2}
   \frac{ \langle ({\cal O}^{\ell_3 \ell_4}_T)^2\rangle-
          \langle {\cal O}^{\ell_3 \ell_4}_T\rangle^2}{
         |\langle{\cal O}^{\ell_3 \ell_4}_T\rangle|^2\,\sigma_{tot}}\,,
\end{eqnarray}
where $\sigma_{tot}=\sigma(p\bar{p}\rightarrow\tilde{\chi}^-_1\tilde{\chi}^0_2)
\,{\cal B}(\tilde{\chi}^0_2\rightarrow\tilde{\chi}^0_1\,\ell^{\prime +}
\ell^{\prime -})$ and
$\langle X\rangle \equiv \int X\,\frac{{\rm d}\sigma_{tot}}{{\rm d}\Phi}\,
{\rm d}\Phi/\sigma_{tot}$ over the total phase space volume $\Phi$.
The numerical factor 2 in the denominator is due to two possible combinations 
of the two final--state leptons; $(e^-,e^+)$ and $(\mu^-,\mu^+)$.\\

Figure~\ref{fig:fig14} exhibits the region of the CP--violating  
phases $\Phi_\mu$ and $\Phi_1$ that is excluded by both the electron EDM 
constraints at 95\% confidence level (shaded region) 
and by the T--odd observable ${\cal O}^{l_3 l_4}_T$ at the 2--$\sigma$ level 
with an integrated luminosity of 20 fb$^{-1}$ (filled circles) and
30 fb$^{-1}$ (open circles) in the scenario ${\cal S}2$.  
Certainly, after incorporating all the systematic errors, the 
covered region should be
reduced. Nevertheless, it might be useful to measure the T--odd observable
at the upgraded Tevatron with an integrated luminosity ${\cal L}$ of the order
of 30 ${\rm fb}^{-1}$ because the electron EDM and  the T--odd TMP
are complementary in constraining the CP--violating phases.

\section{Conclusions}
\label{sec:conclusion}

In this paper, we have investigated in detail the impact of the phases 
$\Phi_\mu$ and $\Phi_1$ on the SUSY tri--lepton signals at the Tevatron 
in the framework of MSSM with general CP phases but without generational 
mixing. The stringent constraints by the electron and neutron EDM
on the CP phases have been also included in the discussion of the
effects of the CP phases.
For the sake of illustration, we have
considered two exemplary scenarios for the relevant SUSY parameters; 
${\cal S}1$ with very heavy first-- and second--generation sfermions
and ${\cal S}2$ with relatively light sfermions but a large $|\mu|$.\\

We have found that in both scenarios the CP--violating phases can have
a significant impact on the production cross section and the partial 
leptonic branching fractions 
of the chargino $\tilde{\chi}^\pm_1$ and neutralino $\tilde{\chi}^0_2$. 
As a result, there may lead to a 
minimum rate of the tri--lepton signal for non--trivial CP phases. 
This implies that one should be careful when interpreting the chargino and
neutralino mass limits derived under the assumption of vanishing phases, 
since the worst case is not (always) covered by just flipping 
the sign of $\mu$; rather it can occur from some non--trivial phases in 
between.\\

The production--decay spin correlations lead to several CP--even 
observables. We have studied the useful kinematical observables such as
the dilepton invariant mass distribution, the angular distributions 
of the final--state leptons, 
and the T--odd (CP--odd) triple momentum products of the initial proton 
momentum and two final lepton momenta.\\

We have found that the the end point of the dilepton invariant mass distribution 
is very sensitive
to the relevant CP--violating phases because of the strong dependence of
the neutralino masses on the phases. Therefore, these distributions can be
very useful in determining the phases once the other real SUSY parameters
are known. The angular distributions of the final--state leptons taking
into account the full spin/angular correlations can differ from the 
non--correlated ones by a few percents. Therefore, it will be
sometimes necessary to consider the fully--correlated distributions
in order to interpret experimental data properly.\\

It turned out to be difficult to investigate the CP--violating phases
directly through the T--odd triple momentum products at the Tevatron
with its upgraded luminosity of about 2 fb$^{-1}$. But, we have 
found that a substantial region of the CP--violating phases may be
explored through the triple momentum products with the luminosity
of about 30 fb$^{-1}$ as proposed for TeV33.

\section*{Acknowledgments}

S.Y.C. and W.Y.S acknowledge financial support of the 1997 Sughak program
of the Korea Research Foundation and M.G. acknowledges Alexander von 
Humboldt Stiftung foundation for financial help. H.S.S. is supported in 
part by the BK21 program.


%


%

%

%

%

\begin{figure}
\begin{center}
\hbox to\textwidth{\hss\epsfig{file=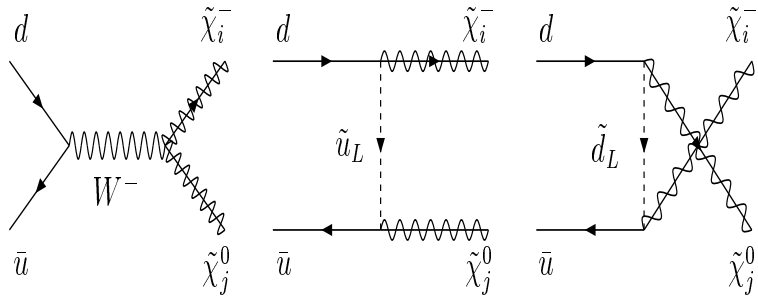,width=10cm,height=4cm}\hss}
\end{center}
\caption{\it Three mechanisms contributing to the parton--level production 
         process  $d\bar{u}\rightarrow \tilde{\chi}^-_i 
          \tilde{\chi}^0_j$; the $s$--channel $W^-$ exchange, the $t$--channel
         $\tilde{u}_L$ exchange and the $u$--channel $\tilde{d}_L$ exchange.}
\label{fig:fig5}
\end{figure}
%

\begin{figure}
\begin{center}
\hbox to\textwidth{\hss\epsfig{file=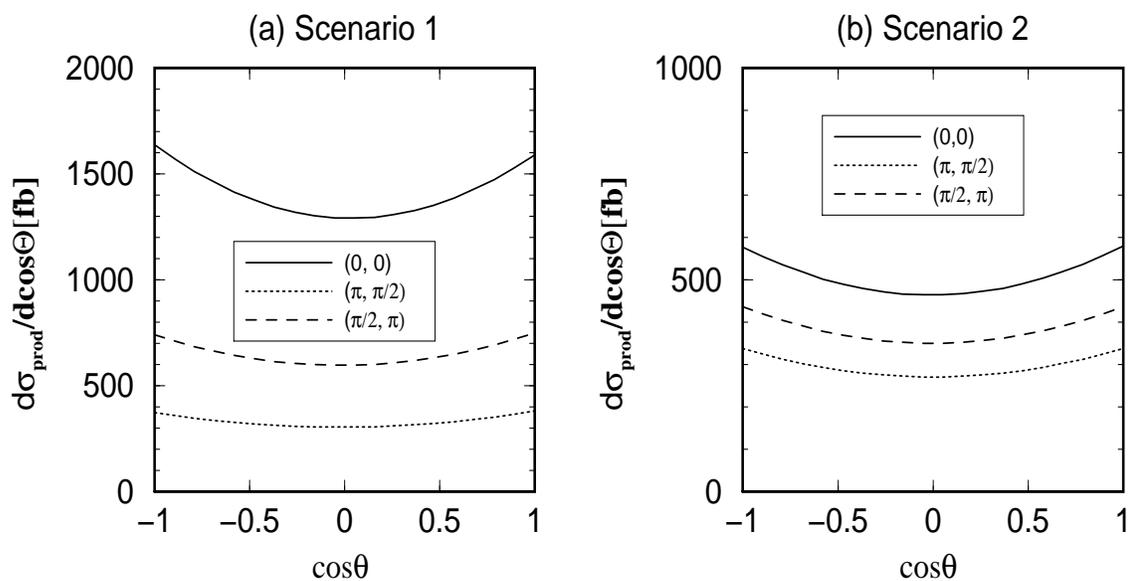,width=15cm,height=8cm}\hss}
\end{center}
\caption{\it The differential production cross section ${\rm d}\sigma_
         {\rm prod} 
         (p\bar{p}\rightarrow\tilde{\chi}^-_1 \tilde{\chi}^0_2+X)/{\rm d}
         \cos\theta$ with respect to the cosine of the scattering angle,
         $\cos\theta$, in (a) ${\cal S}1$ and 
         (b) ${\cal S}2$ for three different sets of $\{\Phi_\mu,\Phi_1\}$.}
\label{fig:fig6}
\end{figure}
%

\begin{figure}
\begin{center}
\hbox to\textwidth{\hss\epsfig{file=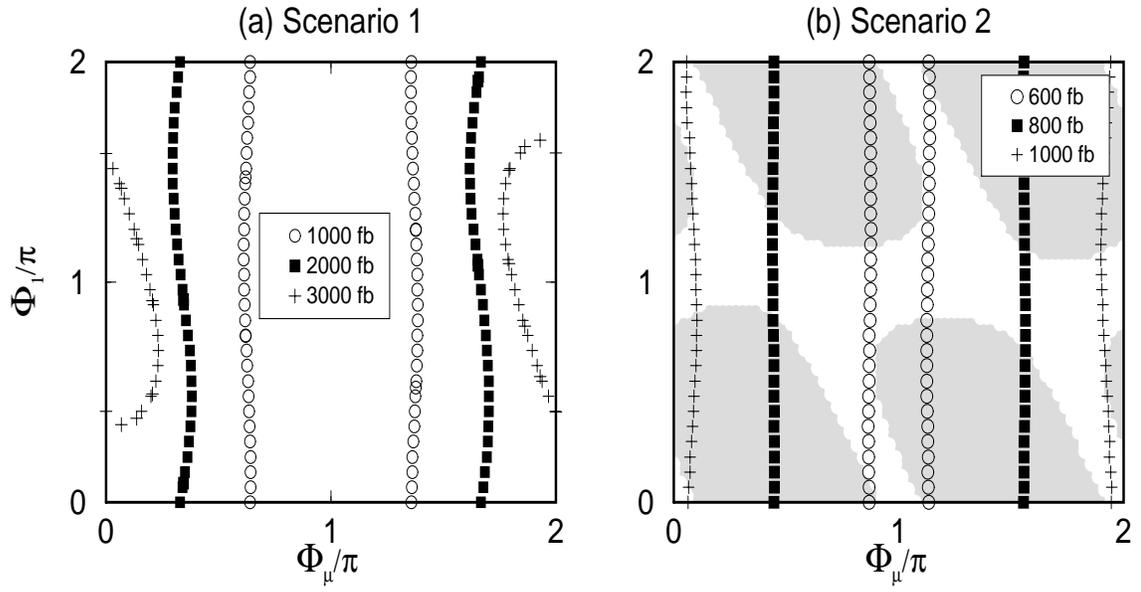,width=15cm,height=8cm}\hss}
\end{center}
\caption{\it The production cross section $\sigma (p\bar{p}\rightarrow
          \tilde{\chi}^-_1\tilde{\chi}^0_2+X)$ on the $\{\Phi_\mu,\Phi_1\}$ 
          plane in (a) the scenario ${\cal S}1$ and (b) the scenario 
          ${\cal S}2$.} 
\label{fig:fig7}
\end{figure}
%

\begin{figure}
\begin{center}
\hbox to\textwidth{\hss\epsfig{file=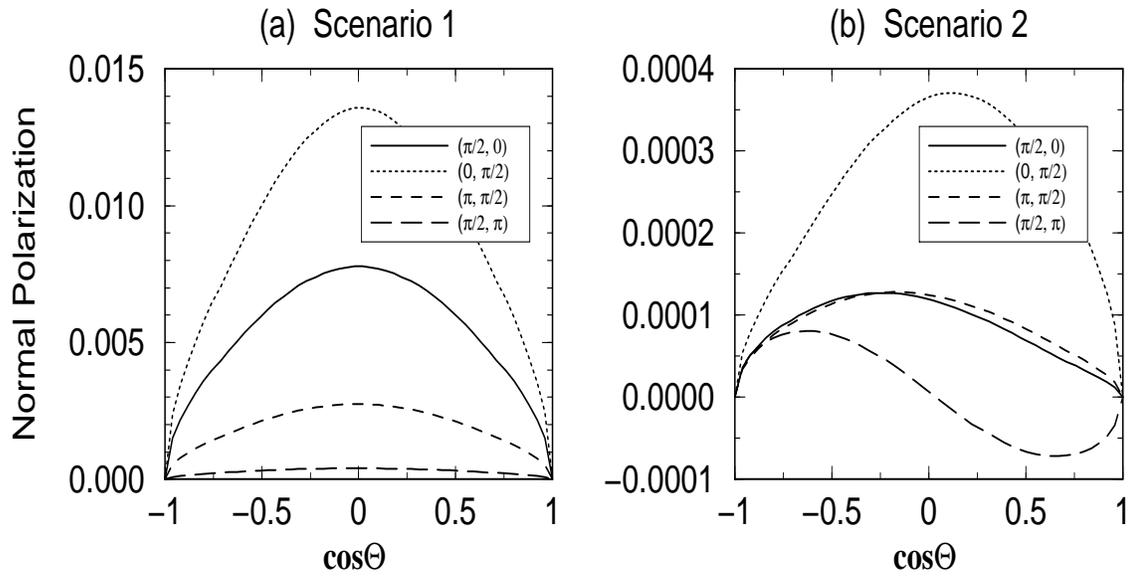,width=15cm,height=8cm}\hss}
\end{center}
\caption{\it The normal polarization component of the neutralino 
         $\tilde{\chi}^0_2$ with respect to the cosine of the
         scattering angle, $\cos\Theta$, 
         in the process $d\bar{u}\rightarrow\tilde{\chi}^-_1\tilde{\chi}^0_2$ 
         for four different combinations of $\{\Phi_\mu,\Phi_1\}$ in 
         (a) ${\cal S}1$ and (b) ${\cal S}2$. The parton--level c.m. energy
         is assumed to be 300 GeV for the sake of illustration.}
\label{fig:fig8}
\end{figure}
%

\begin{figure}
\begin{center}
\hbox to\textwidth{\hss\epsfig{file=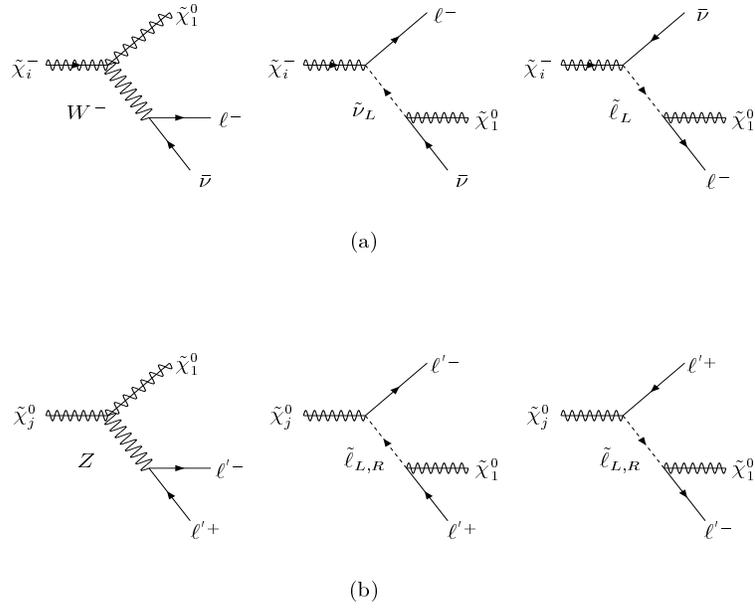,width=10cm,height=8cm}\hss}
\end{center}
\caption{\it The three mechanisms contributing to (a) the chargino decay 
          $\tilde{\chi}^-_i\rightarrow\tilde{\chi}^0_1\,\ell^-\bar{\nu}_\ell$ 
          and (b) the neutralino decay $\tilde{\chi}^0_j\rightarrow
          \tilde{\chi}^0_1\,\ell^{\prime +}\ell^{\prime -}$. 
          In both decays the exchanges of 
          the charged and neutral Higgs bosons are neglected because 
          they involve tiny lepton Yukawa couplings.}
\label{fig:fig9}
\end{figure}
%

\begin{figure}
\begin{center}
\hbox to\textwidth{\hss\epsfig{file=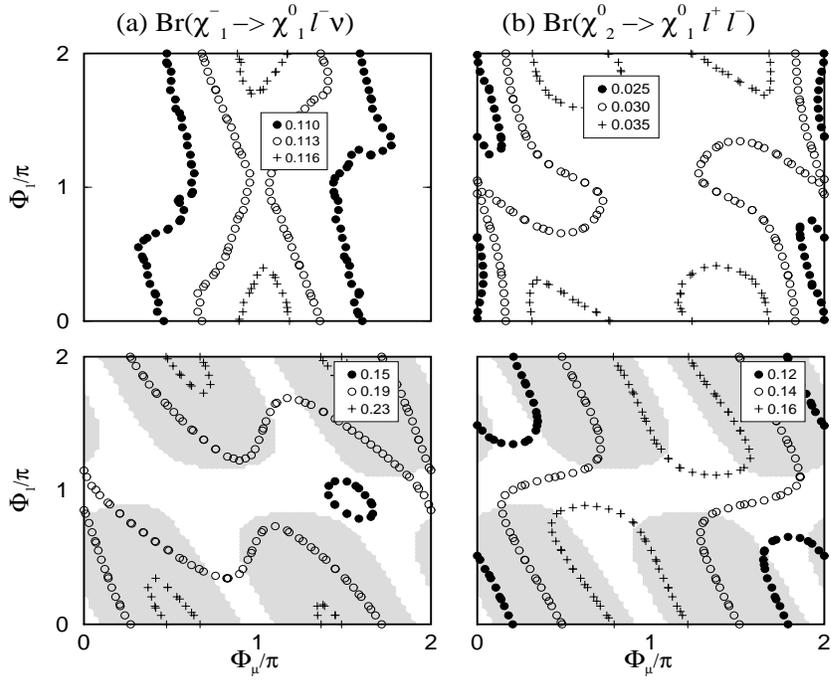,width=11cm,height=9cm}\hss}
\end{center}
\caption{\it The branching fractions ${\cal B}(\tilde{\chi}^-_1\rightarrow
         \tilde{\chi}^0_1\,\ell^-\bar{\nu}_\ell)$ and ${\cal B}(\tilde{\chi}^0_2
         \rightarrow\tilde{\chi}^0_1\,\ell^{\prime +}\ell^{\prime -})$ 
         for $\ell,\, \ell^\prime =e$ or $\mu$ on the $\{\Phi_\mu,\Phi_1\}$ 
         plane in the scenarios ${\cal S}1$ (upper part) and ${\cal S}2$ 
         (lower part).}
\label{fig:fig10}
\end{figure}
%

\begin{figure}
\begin{center}
\hbox to\textwidth{\hss\epsfig{file=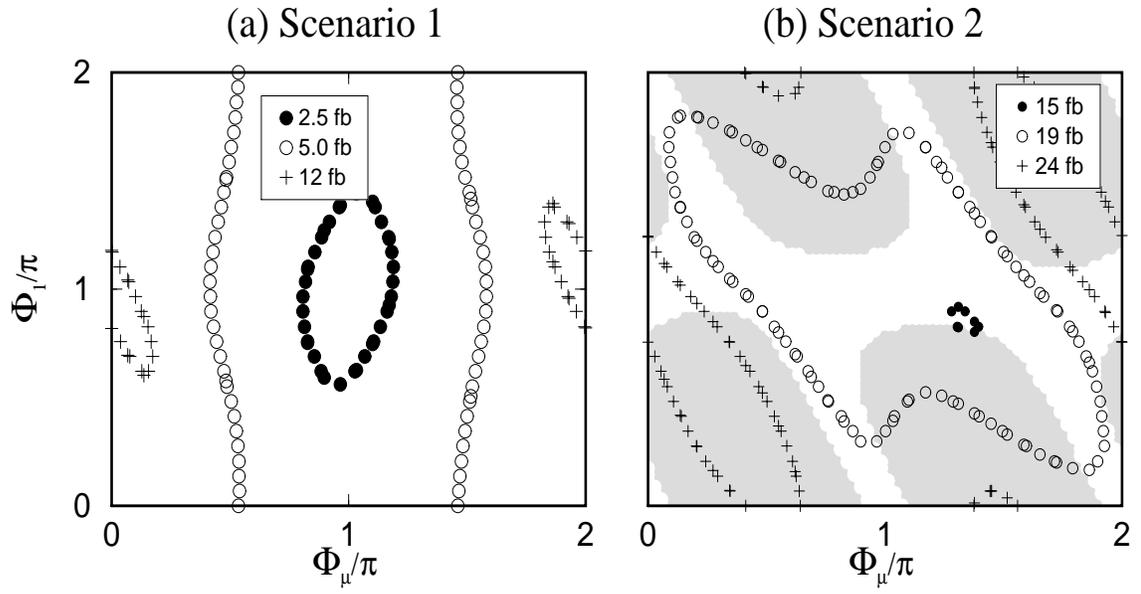,width=15cm,height=8cm}\hss}
\end{center}
\caption{\it The total cross section $\sigma(p\bar{p}\rightarrow 
         3\ell+{\rm X})$ of the tri--lepton signal on the $\{\Phi_\mu,\Phi_1\}$
         plane in the scenarios (a) ${\cal S}1$ and (b) ${\cal S} 2$.}
\label{fig:fig11}
\end{figure}
%

\begin{figure}
\begin{center}
\hbox to\textwidth{\hss\epsfig{file=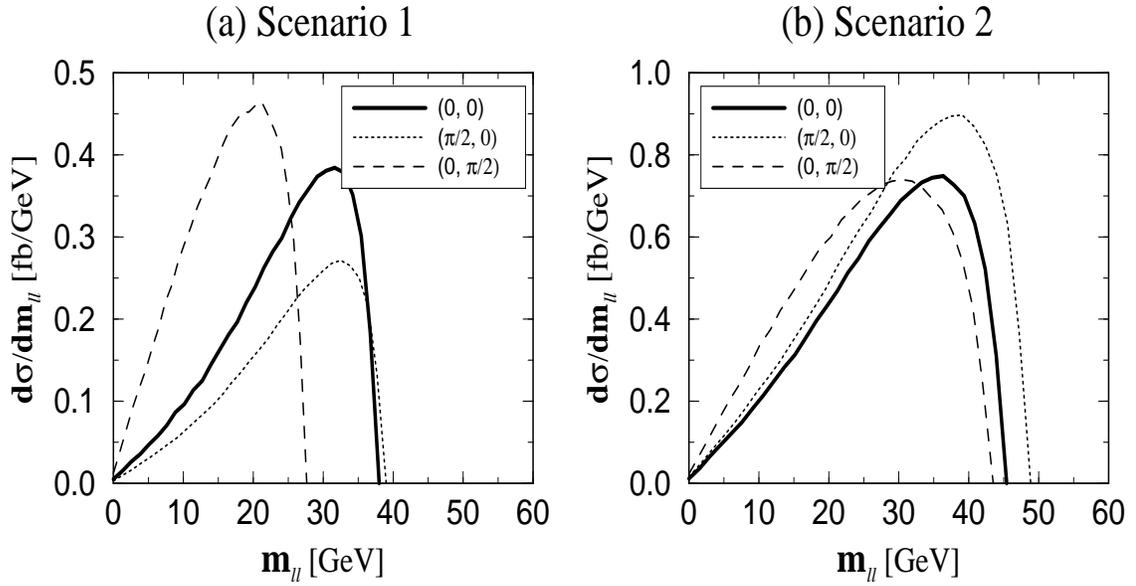,width=15cm,height=8cm}\hss}
\end{center}
\caption{\it The differential correlated cross section ${\rm d}\sigma_{\rm
         tot}/{\rm d}m_{\ell\ell}$ with respect to the invariant mass 
         $m_{\ell\ell}$ of two leptons from the neutralino decay 
         $\tilde{\chi}^0_2\rightarrow\tilde{\chi}^0_1\,\ell^{\prime +}
         \ell^{\prime -}$ for three different sets of 
         $\{\Phi_\mu,\Phi_1\}$ in the scenarios (a) ${\cal S}1$ and 
         (b) ${\cal S}2$.}
\label{fig:fig12}
\end{figure}
%

\begin{figure}
\begin{center}
\hbox to\textwidth{\hss\epsfig{file=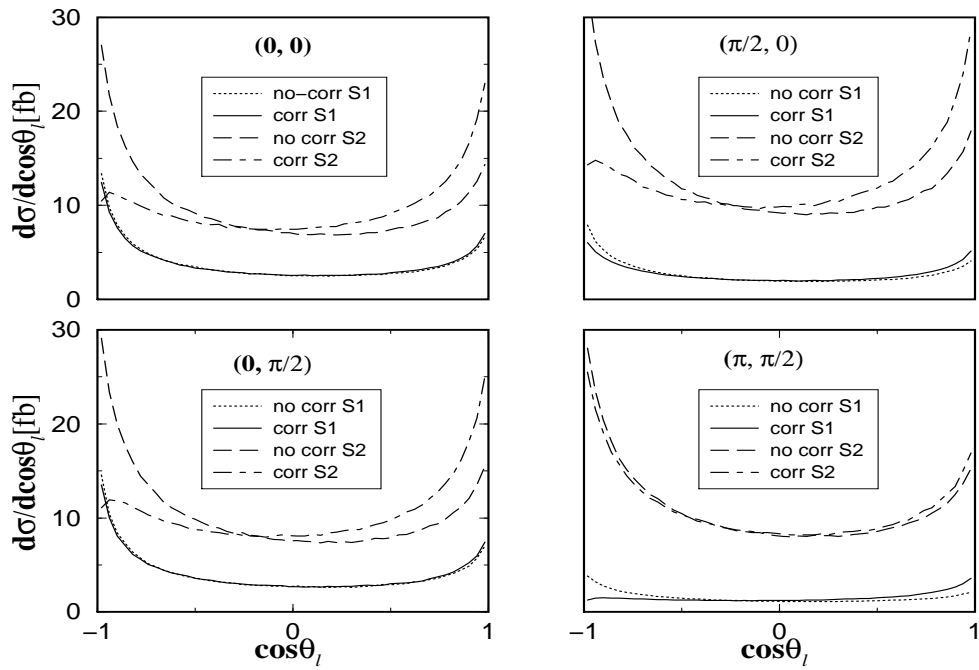,width=13cm,height=9cm}\hss}
\end{center}
\caption{\it The angular distribution of the charged lepton from the lightest
          chargino in the tri--lepton signal for the phases $\{\Phi_\mu,
          \Phi_1\}$; $\{0,0\}$, $\{\pi/2,0\}$, $\{0,\pi/2\}$
          and $\{\pi,\pi/2\}$ in the scenarios ${\cal S}1$ and
          ${\cal S}2$.} 
\label{fig:fig13}
\end{figure}
%

\begin{figure}
 \begin{center}
 \hbox to\textwidth{\hss\epsfig{file=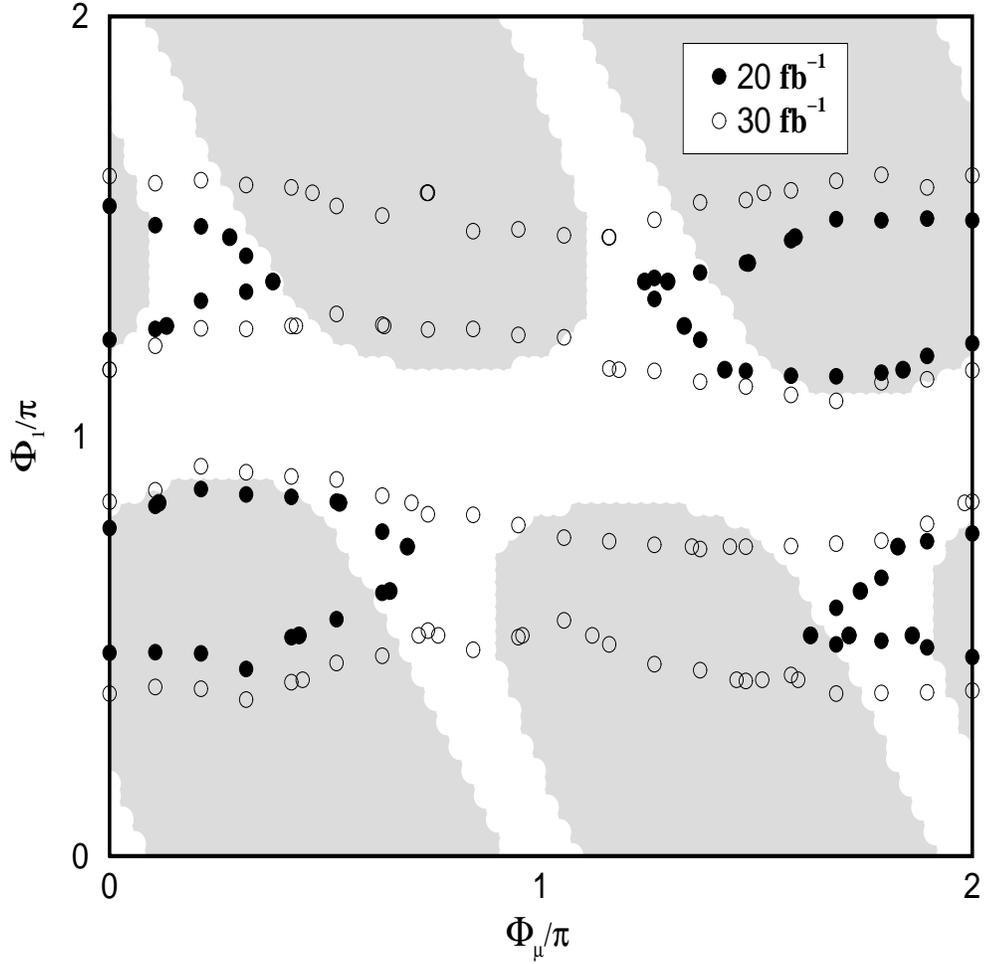,width=13cm,height=13cm}\hss}
 \end{center}
 \caption{\it The region of the CP--violating phases 
           $\{\Phi_\mu,\Phi_1\}$ excluded by the electron EDM measurements 
           (black shadowed region) and probed by the T-odd TMP 
           ${\cal O}^{l_3l_4}_T$ measurements with the integrated luminosity 
           of 20 fb$^{-1}$ (filled circles) and 30 fb$^{-1}$ (open circles) at
           the 2--$\sigma$ level in the scenario ${\cal S}2$.}
 \label{fig:fig14}
\end{figure}

\end{document}